\documentclass[twocolumn]{aastex631}

\AtBeginDocument{}

\usepackage{amssymb}
\usepackage{amsmath}
\usepackage{empheq}
\usepackage{mathrsfs}
\usepackage{wasysym}
\usepackage{fourier-orns}
\usepackage{adforn}

\usepackage{etoolbox}
\usepackage{lmodern}

\newcommand{\be}{\begin{align}}
\newcommand{\ee}{\end{align}}
\newcommand{\G}{\mathcal{G}}
\newcommand{\M}{M_{\star}}

\newcommand{\appropto}{\mathrel{\vcenter{
  \offinterlineskip\halign{\hfil$##$\cr
    \propto\cr\noalign{\kern2pt}\sim\cr\noalign{\kern-2pt}}}}}

\begin{document}

\title{\bf Quasi-Localization of Short-Wavelength Perturbations in Softened Laplace--Lagrange Chains}

\author[0000-0002-7094-7908]{Konstantin Batygin}
\affiliation{Division of Geological and Planetary Sciences, California Institute of Technology, Pasadena, CA 91125, USA}

\author[0000-0002-5486-7828]{Eiichiro Kokubo}
\affiliation{Center for Computational Astrophysics, National Astronomical Observatory of Japan, Osawa, Mitaka, Tokyo 181-8588, Japan}

\begin{abstract}
We examine the long-term evolution of localized eccentricity perturbations in a softened Laplace--Lagrange chain. The secular coupling between mass elements is long-ranged, suggesting that any disturbance imposed at a single radius should disperse rapidly across the chain. We show that this expectation fails once the secular interaction is regularized by a finite softening length $\beta$: while broad eccentricity profiles propagate smoothly, as textbook secular theory anticipates, narrow profiles remain confined to their launching site for millions of secular periods. The origin of this quasi-localization lies in the dispersion relation of the softened chain. The group velocity of eccentricity waves collapses by orders of magnitude on spatial scales below the softening length, so that narrow perturbations project almost entirely onto a nearly flat sector of the band, where little differential precession is available to disperse them. Propagating the initial wavenumber spectrum through the dispersion relation reproduces the retained fraction of angular momentum deficit across the full range of perturbation widths with no adjustable parameters. The confined perturbation is flanked by an apsidally polarized (oppositely-aligned) response, which we show to be the real-space imprint of this spectral projection. The softening length therefore sets a dynamical scale below which eccentricity perturbations spread only on timescales far longer than the secular period.
\end{abstract}


\section{Introduction}

Over the past two decades, the prevailing picture of how solid material is organized within young circumstellar disks has undergone a substantial transformation. The classical view --- in which solids are distributed nearly smoothly across a radially extended power-law profile \citep{1977Ap&SS..51..153W, 1981PThPS..70...35H} --- has yielded to one in which much of the planet-forming inventory is partitioned into discrete annular reservoirs. At extrasolar scales, this restructuring is most directly conveyed by high-resolution ALMA imagery of nearby protoplanetary nebulae, where dust continuum emission is consistently resolved into systems of narrow rings, gaps, and concentric arcs \citep{2015ApJ...808L...3A, 2018ApJ...869L..41A}. Closer to home, isotopic analyses of meteoritic samples have established that the inner Solar System hosted at least two persistently non-mixing reservoirs of solids --- the so-called non-carbonaceous (NC) and carbonaceous (CC) dichotomy \citep{2011E&PSL.311...93W, 2017PNAS..114.6712K} --- whose inferred coexistence over a multi-Myr interval points to the presence of a long-lived dynamical barrier between adjacent radial zones. Moreover, the architecture of the terrestrial planets themselves --- in particular, the small masses of Mars and Mercury relative to Venus and Earth --- has long resisted a smooth-disk explanation, and is best reproduced by initial conditions in which terrestrial precursors are confined to a narrow annulus near $\sim 1$\,au \citep{2009ApJ...703.1131H, 2011Natur.475..206W, 2022NatAs...6..357I, 2025arXiv250405667K, 2026arXiv260104032G}. Taken together, these strands of evidence point to a planet formation environment whose elementary unit is a narrow ring rather than a continuous distribution.

This evolving picture brings with it a host of new dynamical questions. In what follows, we take up only one small facet of this broader landscape: how, within a disk whose mass is concentrated into discrete annuli, gravitational disturbances --- and, more generally, the system's angular momentum deficit (AMD; \citealt{1997A&A...317L..75L}) --- are redistributed across radial scales -- a question that must be answered before ring-chain secular models can be brought to bear on the longevity or interaction of the observed annular structures. Even this narrow question admits a useful conceptual decomposition.

At a broad level, gravitational dynamics in a many-body system splits naturally into two complementary channels. The granularity of the mass distribution supports discrete encounters whose cumulative effect drives standard two-body relaxation \citep{1942psd..book.....C}. Although this channel is itself long-ranged -- the net deflection is dominated by the accumulation of many weak, distant encounters, as reflected in the Coulomb logarithm -- its defining signature lies elsewhere: encounters transpire between bodies at particular orbital phases, and the associated exchanges of energy diffuse the participants' semi-major axes. The smooth component of the underlying potential, by contrast, mediates collective phenomena, in which large numbers of mass elements respond coherently to a self-consistent mean field. This decomposition is mirrored in plasma physics, where the dynamical hierarchy of a charged $N$-body system is organized by the splitting between the collisional (Boltzmann) and collective (Vlasov) channels -- a distinction drawn not by the range of the interaction, but by the collision term that granularity introduces.

In what follows, we focus exclusively on the collective channel: that is, on the smooth gravitational response of a chain of mutually-interacting eccentric annuli, in which the discreteness of individual particles is averaged over. Notably, the secular framework we employ is collisionless not by approximation but by construction: phase-averaging removes the mean anomalies from the problem at the outset, the conjugate actions -- the semi-major axes -- are correspondingly frozen, and the resulting model is the exactly symplectic Hamiltonian flow of a smooth averaged potential, containing no collision integral at any order. The resulting Hamiltonian therefore describes the evolution of an exactly orbit-averaged mean field, retaining only the collective response of the averaged gravitational potential. The residual discreteness of a finite collection of wires can, in principle, still drive a slow, phase-averaged redistribution of the angular momentum deficit in the manner of resonant relaxation \citep{1996NewA....1..149R}; within the linear Laplace--Lagrange theory adopted here, however, the dynamics are purely those of the collective response.

The canonical analytic framework for the collective dynamics of such ring chains is provided by secular perturbation theory, in which phase-averaging over orbital periods reduces each particle to a wire of mass distributed along a Keplerian ellipse, and the gravitational interaction between particles to that between mass-weighted closed curves \citep{1999ssd..book.....M}. The approach has a long and rich pedigree in celestial mechanics. Classical applications include the apse-aligned precession of the Uranian $\epsilon$~ring, where the role of self-gravity in maintaining a coherently-precessing narrow annulus was first proposed by \citet{1979AJ.....84.1638G} and quantitatively resolved by \citet{1983AJ.....88.1560B}; the dynamical assembly of the regular satellite systems of the giant planets from primordial debris disks \citep{2012Icar..219..737M}; and the secular evolution of the Kuiper belt under the gravitational influence of the giant planets, modeled as a system of softened mutually-interacting rings \citep{2003ApJ...595..531H}. Closely related formulations of the same wire-based formalism have, in parallel, been used to study the secular dynamics of self-gravitating eccentric stellar disks around supermassive black holes \citep{2009MNRAS.394.1085T, 2019PhRvL.123b1103T} and the vertical structure of self-gravitating debris disks \citep{2025MNRAS.543.3123S}. Beyond its utility as a numerical tool, the secular-ring formalism admits a mathematical connection to the time-dependent Schr\"odinger equation \citep{2018MNRAS.475.5070B, MeltonBatygin}.

A central technical ingredient in any practical implementation of secular-ring theory is the manner in which the singular near-neighbor interaction is regularized. Adjacent wires whose semi-major axes differ by less than their physical thicknesses --- whether interpreted as a vertical disk scale-height or velocity-dispersion floor --- couple through coefficients that diverge as the inverse square of their fractional separation as the orbits approach exact crossing. The customary remedy is to introduce a softening parameter, hereafter denoted $\beta$, that smoothly excises this singularity from the orbit-averaged potential, and this is the approach we shall adopt herein as well.

The key result of this paper can be stated compactly. Softening endows the secular interaction with a dominant coupling profile of finite radial width. The Fourier transform of this profile -- the dispersion relation of the chain -- accordingly develops a nearly flat short-wavelength sector, across which the group velocity of eccentricity disturbances collapses by orders of magnitude. Perturbations narrower than the softening length project almost entirely onto this flat sector and, with very little differential precession available to disperse them, persist as coherent, spatially confined structures for times exceeding the secular period by many orders of magnitude. The body of the paper establishes this chain of reasoning quantitatively, and tests its robustness against resolution, chain inhomogeneity, and the choice of softening prescription.

We remark that our results concern the softened secular formalism itself, not any particular disk. The confinement we document is a property of the regularized interaction operator; whether a real disk would preserve the same structure depends on processes absent from the present model. In practice, ring-chain secular calculations of the annular architectures described above will, if they resolve structure below the softening length, preserve that structure for times that can exceed the age of the system, and this persistence should not be mistaken for a dynamical prediction. The short-wavelength flattening of the secular band that underlies it has antecedents in the continuum theory of softened discs \citep{2001AJ....121.1776T, 2003ApJ...595..531H}; the contribution of the present work is the quantitative initial-value problem -- how much of a perturbation is retained as a function of its width, why, for how long, and how the answers depend on the softening prescription.

The remainder of this paper is organized as follows. In Section~2, we describe the implementation of our softened Laplace--Lagrange model and derive the explicit form of the regularized secular coupling coefficients. Section~3 presents a series of semi-analytical experiments that probe the propagation of disturbances within the chain across a range of perturbation widths, identifies a regime of quasi-localization at short wavelengths, traces it to the dispersion relation of the softened chain, which renders the quasi-localization quantitative, and shows that the polarized morphology of the confined response follows from the same spectral structure. Section~4 summarizes our findings, and returns to the physical meaning of the softening parameter $\beta$.

\section{Softened Laplace--Lagrange Model}\label{sec:setup}

We consider a chain of $N$ coplanar, gravitating wires of mass $m_j$ orbiting a dominant central body of mass $\M$ at semi-major axes $a_j$. Each wire is described in terms of the complex eccentricity vector
\begin{align}
z_j \;=\; h_j + i\,k_j \;=\; e_j\,e^{i\varpi_j},
\label{eqn:zdef}
\end{align}
where $h_j = e_j\cos\varpi_j$ and $k_j = e_j\sin\varpi_j$ are coordinates closely related to the standard Poincar\'e variables, and $\varpi_j$ denotes the longitude of pericenter. Throughout the analysis, we restrict ourselves to the leading non-trivial order in eccentricity, so that the orbit-averaged Hamiltonian governing the long-term evolution of the chain takes the standard quadratic form
\begin{align}
\mathcal{H}_{\rm sec}
= \frac{1}{2}\sum_{j} \Lambda_{j}\,A_{jj}\,|z_j|^{2}
+ \sum_{j< k} \Lambda_{j}\,A_{jk}\,\mathrm{Re}\bigl(z_j\,z_k^{\ast}\bigr),
\label{eqn:Hsec}
\end{align}
in which $\Lambda_{j} \equiv m_{j}\,n_{j}\,a_{j}^{2}$ is the circular angular momentum of wire $j$, and the matrix $\mathbf{A}$ encapsulates the gravitational couplings, depending only on the masses and the semi-major axes (note that $\Lambda_{j}\,A_{jk} = \Lambda_{k}\,A_{kj}$, so that the quadratic form is symmetric).

Treating $z_{j}$ and $z_{j}^{\ast}$ as independent variables in the complex-coordinate Hamiltonian sense \citep{1966RvMP...38...36S}, with the $\Lambda$-weighted bracket $\dot{z}_{j} = (2\,i/\Lambda_{j})\,\partial \mathcal{H}_{\rm sec}/\partial z_{j}^{\ast}$, the associated equations of motion are linear,
\begin{align}
\frac{dz_j}{dt} \;=\; i\sum_{k} A_{jk}\,z_{k},
\label{eqn:LLeom}
\end{align}
and admit a complete eigenmode decomposition with real eigenfrequencies $\{\lambda_n\}$, such that
\begin{align}
z_j(t) &\;=\; \sum_{n} V_{jn}\,c_n\,e^{i\lambda_n t}, \nonumber \\
c_n &\;=\; \bigl(\mathbf{V}^{-1}\,\mathbf{z}(0)\bigr)_n,
\label{eqn:zsol}
\end{align}
where $\mathbf{V}$ is the matrix of eigenvectors of $\mathbf{A}$.\footnote{Although $\mathbf{A}$ itself is not symmetric in the Euclidean sense, the $\Lambda$-weighted matrix $\Lambda_{j} A_{jk}$ appearing in equation~(\ref{eqn:Hsec}) is, so that the spectrum of $\mathbf{A}$ is real and its eigenvectors are mutually orthogonal under the corresponding weighted inner product \citep{1999ssd..book.....M}.} At its core, then, second-order Laplace--Lagrange theory reduces the long-term dynamics of an annulus of mutually-interacting wires to a linear-algebraic eigenvalue problem, with the entirety of the gravitational interaction encoded in the matrix $\mathbf{A}$.

Following the standard textbook derivation \citep{1999ssd..book.....M}, the off-diagonal entries of $\mathbf{A}$ encode pairwise coupling between distinct wires, while the diagonal entries collect the apsidal precession contributed by all other wires of the chain:
\begin{align}
A_{jk} &= n_j\,\frac{m_k}{\M}\,\bar{\alpha}_{jk}\,C_2\bigl(\alpha_{jk},\beta\bigr), \quad j\neq k, \nonumber \\
A_{jj} &= n_j \sum_{k\neq j}\frac{m_k}{\M}\,\bar{\alpha}_{jk}\,C_1\bigl(\alpha_{jk},\beta\bigr),
\label{eqn:Aentries}
\end{align}
where $n_j = \sqrt{\G \M / a_j^3}$ is the Keplerian mean motion of the $j$-th wire, $\alpha_{jk} = \min(a_j, a_k)/\max(a_j, a_k)$, and $\bar{\alpha}_{jk} = a_j/a_k$ when $a_j < a_k$ and unity otherwise. The dimensionless coefficients $C_1$ and $C_2$ are dictated entirely by the orbit-averaged geometry of the pair of interacting wires.

\subsection{Softening of the Laplace Coefficients}

In the unsoftened limit, the coefficients $C_1$ and $C_2$ are constructed from $\alpha$-derivatives of the classical Laplace coefficient $b_{1/2}^{(m)}(\alpha)$, and both diverge as $(1-\alpha)^{-2}$ as $\alpha \to 1$ (the Laplace coefficient itself diverges only logarithmically; its derivatives diverge algebraically). Because the phase-averaging procedure that underlies Laplace--Lagrange theory implicitly treats each wire as a continuous closed curve, the divergence at $\alpha=1$ is unphysical: it reflects only the incidental fact that the bare gravitational potential between two coplanar Keplerian ellipses is unbounded at the instant of formal orbit crossing. In any physically motivated implementation, this singularity must be regularized by an external scale that captures the finite vertical thickness, internal velocity dispersion, or finite particle size of the system under consideration.

A range of softening prescriptions has been employed in the literature. \citet{1981Icar...47..234W} introduced an effective ``thickness'' to capture the vertical extent of an axisymmetric disk; \citet{1998AJ....116.2015T} formulated softened secular dynamics for self-gravitating eccentric stellar disks around supermassive black holes; \citet{2003ApJ...595..531H} adopted a softened Laplace--Lagrange theory in his treatment of the secular Kuiper belt; and \citet{2009MNRAS.394.1085T} derived a Gauss-style softened secular Hamiltonian valid at arbitrary eccentricity and inclination. A useful synthesis and generalization of these approaches is provided by \citet{2019MNRAS.489.4176S}, who clarified the relationship between several of the most commonly-employed softening prescriptions and discussed their suitability across a range of astrophysical settings.

The approach we adopt here is among the most direct: we retain the textbook expressions for $C_1$ and $C_2$ \citep{1999ssd..book.....M}, written explicitly in terms of $\alpha$-derivatives of $b_{1/2}^{(m)}(\alpha)$, and regularize the singularity by adding a Plummer-type softening parameter $\beta$ inside the integrand. Concretely, we replace the standard Laplace coefficient with the softened form
\begin{align}
B_{1/2}^{(m)}(\alpha,\beta)
\;=\; \frac{2}{\pi}\!\int_{0}^{\pi}\!\!
\frac{\cos(m\,\psi)\,d\psi}
{\bigl(1+\alpha^{2}+\beta^{2}-2\alpha\cos\psi\bigr)^{1/2}},
\label{eqn:bsoft}
\end{align}
which reduces to the textbook expression $b_{1/2}^{(m)}(\alpha)$ in the $\beta \to 0$ limit; we use the upper-case symbol throughout to distinguish the softened coefficients from their classical counterparts. In this notation, the two secular coupling coefficients become
\begin{align}
C_1(\alpha,\beta) &= \frac{1}{4}\Bigl[\, 2\alpha\,\partial_{\alpha} B_{1/2}^{(0)} + \alpha^{2}\,\partial_{\alpha}^{2} B_{1/2}^{(0)} \Bigr], \nonumber \\[2pt]
C_2(\alpha,\beta) &= \frac{1}{4}\Bigl[\, 2\,B_{1/2}^{(1)} - 2\alpha\,\partial_{\alpha} B_{1/2}^{(1)} - \alpha^{2}\,\partial_{\alpha}^{2} B_{1/2}^{(1)} \Bigr],
\label{eqn:C12}
\end{align}
with the $\alpha$-derivatives of equation~(\ref{eqn:bsoft}) evaluated analytically under the integral sign and the resulting trigonometric integrals computed by quadrature. Because $\beta$ is retained inside the radical, the integrand is everywhere bounded, including at $\alpha=1$.

Although the classical recursion identities relating higher-order Laplace coefficients to derivatives of $b_{1/2}^{(0,1)}$ do not carry over unchanged to the softened case, equivalent representations exist: \citet{2019MNRAS.489.4176S} express the same coefficients through softened Laplace coefficients of higher order (their Table~1) and provide closed forms in terms of complete elliptic integrals (their Appendix~C3). Direct differentiation under the integral sign is therefore a matter of convenience rather than necessity, and we adopt it because it yields the explicit asymptotic forms of Appendix~\ref{app:A} directly. In the classification of \citet{2019MNRAS.489.4176S}, equations~(\ref{eqn:bsoft})--(\ref{eqn:C12}) constitute the prescription of \citet{1998AJ....116.2015T}: they are the exact secular coefficients of the pairwise potential $\G m'/\sqrt{\mathcal{D}^{2} + \beta^{2}r'^{2}}$, softened by a fixed fraction $\beta$ of the instantaneous radius $r'$ of the outer wire -- equivalently, for radially non-overlapping orbits, of the larger of the two instantaneous radii (\citealt{2019MNRAS.489.4176S}, Table~B1). The $\delta$-function terms that \citet{2019MNRAS.489.4176S} find for this prescription in the continuum limit (their Section~6.1) arise from the kink of the softening function at coincident radii.

For pairs of distinct wires ($\alpha \neq 1$) they vanish identically, as those authors point out, but the kink itself survives as a cusp of the coefficients at small separation (Appendix~\ref{app:A}), and it is this cusp that sets the residual dispersion of Section~\ref{sec:dispersion}. Relatedly, the precession rates that this prescription yields for a continuous disc, when its coefficients are used uncorrected as they are here, do not converge to the razor-thin values as the softening is removed (\citealt{2019MNRAS.489.4176S}, Sections~3.1 and~6.1; the corrected continuum form of the instantaneous-maximum prescription in their Section~6.1 recovers the razor-thin limit once its distributional terms are included), whereas those of \citet{2003ApJ...595..531H} do; this is immaterial for what follows, since the confinement we study is a finite-softening property of the operator and is reproduced under the latter prescription as well (Appendix~\ref{app:C}). We verify the correspondence against a direct orbit-average of the softened disturbing function in Appendix~\ref{app:verification}, where all three quadratic coefficients of the pairwise interaction are recovered over two decades in the fractional separation $\lambda$ for two softening families, $\beta = 10\lambda$ and $100\lambda$.

\subsection{Numerical Setup}\label{sec:setup_numerical}

For definiteness, all of the semi-analytical experiments described in Section~\ref{sec:experiments} below adopt a single, fixed parameter set. The chain consists of $N = 1000$ wires distributed in geometric succession in semi-major axis,
\begin{align}
a_{j} \;=\; a_{0}\,(1 + \lambda)^{\,j},
\qquad j = 1,\ldots,N,
\label{eqn:aj_geom}
\end{align}
with $a_{0} = 0.95$ and a wire-spacing parameter $\lambda = 10^{-4}$, so that the chain spans the semi-major-axis range $a \in [0.95,\,1.05]$ and the corresponding fractional radial extent of the annulus is $\Delta a/a \simeq 0.10$. The softening is fixed at $\beta = 100\,\lambda = 10^{-2}$. The mass distribution along the chain is taken to be $m_{j} = m_{0}\,a_{j}$. Because the wire spacing is itself geometric, with $\Delta a_{j} \equiv a_{j+1} - a_{j} = \lambda\,a_{j}$, the wire mass per annular area at radius $a_{j}$ -- the local surface density -- evaluates to
\begin{align}
\Sigma(a) \;=\; \frac{m_{j}}{2\pi\,a_{j}\,\Delta a_{j}} \;=\; \frac{m_{0}\,a_{j}}{2\pi\,a_{j}\,(\lambda\,a_{j})} \propto\; \frac{1}{a}.
\label{eqn:Sigma_profile}
\end{align}
The per-wire normalization $m_{0}$ is chosen so that the total mass of the annulus, $M_{\rm disk} \equiv \sum_{j=1}^{N} m_{j} \simeq 5\times 10^{-7}\,\M$, is small compared with the central body. Because the mass, mean motion, and coupling strength of this chain all vary with radius, we refer to it below as the \emph{graded} chain, to distinguish it from the index-homogeneous idealization introduced in Section~\ref{sec:dispersion}.

Whereas $\beta$ can be motivated on physical grounds -- as a disk aspect ratio or a velocity-dispersion floor -- the wire spacing is a purely numerical parameter, and if the aim were to emulate the secular dynamics of a smooth disk, it would have to be chosen small enough for the chain's precession rates to converge to their continuum values. \citet{2019MNRAS.489.4176S} (their Section~6.3 and Figure~9) find that faithful recovery of the smooth-disk precession rate requires a local spacing $\Delta a/a \lesssim 0.1\,\varsigma^{2}$ in terms of their softening parameter $\varsigma$, so that the number of rings must scale as $N \gtrsim 10\,\varsigma^{-2}$ -- for $\beta = 10^{-2}$, a spacing of order $10^{-5}$, an order of magnitude finer than the $\lambda = 10^{-4}$ adopted here. Our chain is therefore not a converged representation of a smooth disk in that sense. What the experiments below establish, however, is a property of the softened chain that is controlled by the ratio $\beta/\lambda$ rather than by either parameter separately, so that as $\lambda \to 0$ at fixed $\beta$ the fate of a perturbation of fixed physical width is dictated by whether it is wider or narrower than the softening length. This expectation is borne out directly: at fixed physical softening, the retained fraction of a perturbation of fixed physical width is constant to $2.4\%$ across a factor of four in resolution, and a single-wire perturbation on the graded chain shows no measurable decay through $3\times10^{7}\,T_{\rm sec}$ at $N = 500$, $1000$, and $2000$ alike, whether the chains are compared in their own secular time or at fixed physical time -- although, as discussed in Appendix~\ref{app:C}, a single-wire perturbation is a grid-scale rather than a fixed physical initial condition, so this establishes the absence of leakage at each resolution rather than the convergence of one fixed initial-value problem.

The corresponding secular timescale is set by the diagonal entries of the Laplace--Lagrange matrix, which scale schematically as $|A_{jj}| \sim n_{j}\, (M_{\rm disk}^{\rm eff}/\M)\,\langle C_{1}\rangle$, where $M_{\rm disk}^{\rm eff}$ is the mass of wires within roughly one softening length of the $j^{\rm th}$ wire and $\langle C_{1}\rangle$ is the typical softened coupling coefficient at those separations (equation~\ref{eqn:C12}). Because the dimensionless coefficients $C_1, C_2$ depend only on $\lambda$ and on the dimensionless ratio $\beta/\lambda$, the secular timescale of any wire obeys $T_{\rm sec}/T_{\rm orb} \propto \M/M_{\rm disk}$ at fixed $\lambda$ and $\beta/\lambda$. For the parameter set adopted here, direct diagonalization of $\mathbf{A}$ yields a median eigenmode period $T_{\rm sec} \equiv \mathrm{median}_{n}(2\pi/|\lambda_{n}|) \simeq 2.9\times 10^{4}$ orbital periods at $a=1$, which serves as a convenient precession-based unit of time for the experiments described in Section~\ref{sec:experiments}. We emphasize that $T_{\rm sec}$ is a clock rather than a transport rate: it measures the typical apsidal precession period of the chain's modes, and is dominated by their common -- dynamically inert, but physically real -- precession. Where transport is at issue, the relevant scale is set instead by the group velocity (Section~\ref{sec:dispersion}); for reference, the fastest wave packets the chain supports cross the full annulus in $\simeq 0.3\,T_{\rm sec}$, so a confinement time of $10^{6}\,T_{\rm sec}$ corresponds to $\sim\!3\times10^{6}$ crossing times of the mobile branch. As emphasized in that section, the absolute mass scale enters \emph{only} through the timescale $T_{\rm sec}$; the eccentricity dynamics themselves are strictly scale-free.

\section{Semi-Analytical Experiments}\label{sec:experiments}

The semi-analytical experiments described in this section address a single question: how does an isolated, narrow eccentricity perturbation imposed at one location within the chain propagate through the remainder of the system over many secular periods? In every experiment, we initialize the chain by setting all longitudes of pericenter to a common value, $\varpi_{j} = 0$ for $j = 1, \ldots, N$, and superimpose on the otherwise-circular state a localized Gaussian perturbation in the eccentricity profile,
\begin{align}
e_{j}(t = 0) \;=\; e_{\rm peak}\,\exp\!\left[\, -\,\frac{(j - j_{c})^{2}}{2\,\sigma^{2}}\, \right],
\label{eqn:gauss_ic}
\end{align}
centered on the central wire of the annulus ($j_{c} = N/2$). The width $\sigma$, measured in units of wire index, is the single free parameter that we vary across experiments.

A useful observation at this stage is that the secular equations of motion (\ref{eqn:LLeom}) are linear and homogeneous in the eccentricity vector $\mathbf{z}$. As a consequence, any solution $\mathbf{z}(t)$ obtained for a given peak amplitude $e_{\rm peak}$ may be multiplied by an arbitrary constant factor without disturbing the equations themselves, and the relative morphology of the response everywhere in the chain is independent of the absolute amplitude of the initial perturbation. The dimensionless width $\sigma$ accordingly remains the only free parameter of the initial condition.\footnote{The propagation depends, in addition, on the fixed parameters of the chain itself -- most importantly the ratio $\beta/\lambda$.} We note, further, that the confinement lifetimes quoted in Section~\ref{sec:dispersion} pertain to this strict linear limit: at finite amplitude, fourth-order secular terms shift the mode frequencies by relative corrections that are quadratic in the eccentricity, with a coefficient that depends on the softening (for closely spaced softened wires, the relevant expansion parameter is the eccentricity difference measured in units of $\beta$ rather than the eccentricity itself), and these corrections can modify the coherence and confinement of a confined perturbation; we have not derived the corresponding timescale for this chain.

\subsection{Eccentricity Propagation and the Emergence of Quasi-Localization}\label{sec:propagation}

\begin{figure*}
\includegraphics[width=\textwidth]{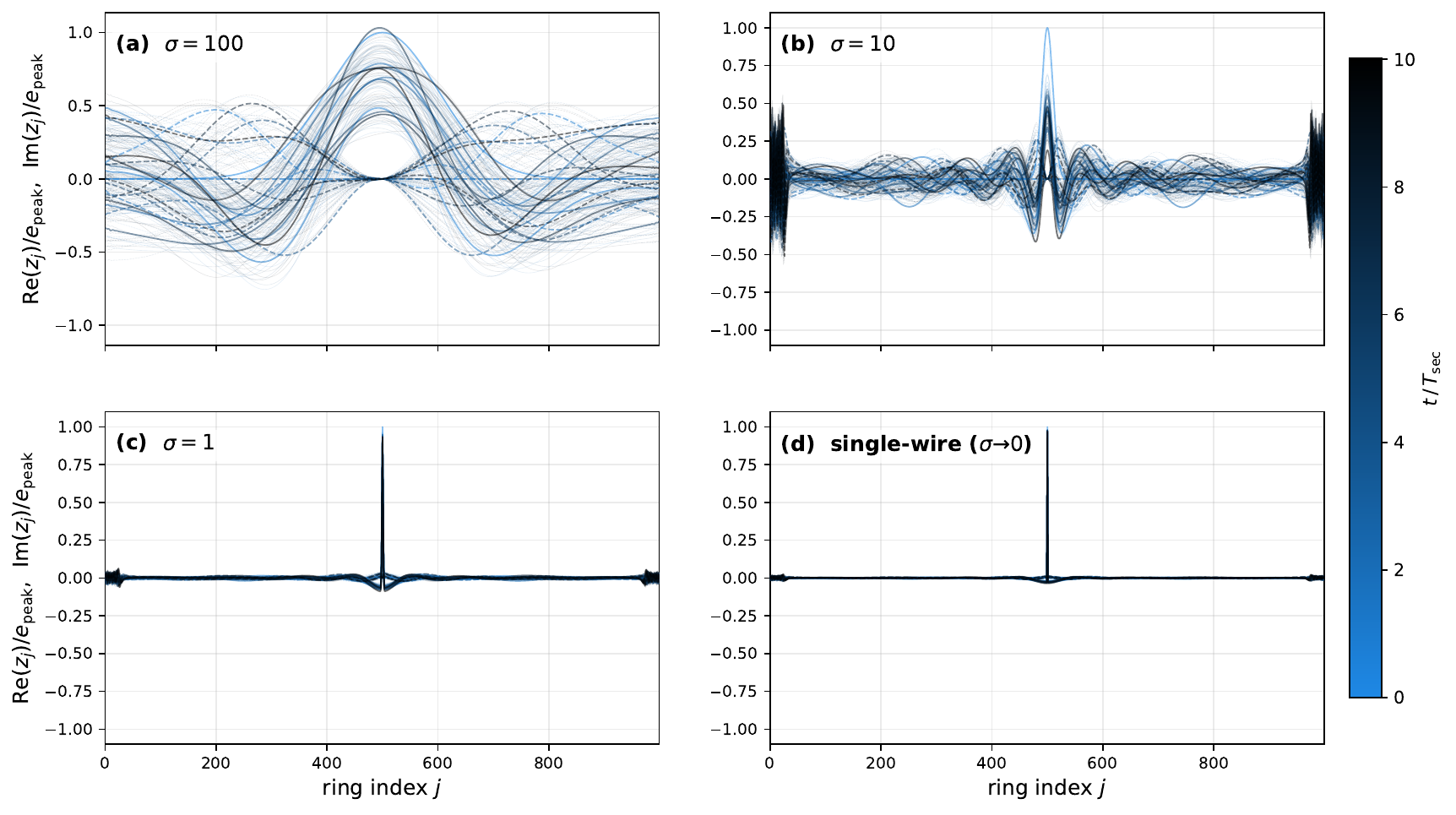}
\centering
\caption{Time evolution of the complex eccentricity vector $z_{j} = h_{j} + i k_{j}$ along the chain, for four Gaussian initial conditions of decreasing width $\sigma$. Each panel shows $\mathrm{Re}(z_{j})/e_{\rm peak}$ (solid) and $\mathrm{Im}(z_{j})/e_{\rm peak}$ (dashed) as functions of the ring index $j$, evaluated in a frame co-rotating with the apsidal line of the central wire $j_{c}$, where $e_{\rm peak} \equiv |z_{j_{c}}(0)|$ is the peak eccentricity of each panel's initial condition, so that the central wire starts at unity in every panel. The pale background traces show $100$ time snapshots uniformly distributed over the interval $t \in [0, 10\,T_{\rm sec}]$, while the darker overlaid curves correspond to integer-multiple snapshots at $t/T_{\rm sec} = 0, 1, \ldots, 10$. The four panels correspond to (a) $\sigma = 100$ (width comparable to the softening length, $\beta/\lambda$), (b) $\sigma = 10$, (c) $\sigma = 1$, and (d) a single-wire $\delta$-function perturbation ($\sigma \rightarrow 0$). Panel (a) exhibits a smoothly-propagating, wave-like response that disperses across the chain, whereas panels (c) and (d) show that perturbations narrower than an intrinsic length scale of the chain do not propagate appreciably over the integration window, remaining trapped at the launching site.}
\label{fig:propagation}
\end{figure*}

Evolving $z$ over many secular periods, we examine the temporal evolution of the chain under the family of Gaussian initial conditions, with the width $\sigma$ swept across a range of physically interesting values. The principal results are summarized in Figure~\ref{fig:propagation}, which shows $\mathrm{Re}(z_{j})/e_{\rm peak}$ and $\mathrm{Im}(z_{j})/e_{\rm peak}$ along the chain at a sequence of eleven time snapshots between $t = 0$ and $t = 10\,T_{\rm sec}$. Because the linear secular evolution conserves the total AMD ($\tfrac{1}{2}\sum_{j}\Lambda_{j}|z_{j}|^{2}$ at the quadratic order retained here) while continually winding the apsidal phases, we display the eccentricity vector in a frame that co-rotates with the apsidal line of the central wire. More precisely, the quantity shown in Figure~\ref{fig:propagation} is the de-rotated eccentricity vector $\tilde z_{j}(t) \equiv z_{j}(t)\,e^{-i\,\varpi_{j_{c}}(t)}$, in which the global apsidal precession of the central wire has been absorbed into the frame; this rotation leaves the relative morphology of the response across the chain unchanged while removing the trivial overall winding. The four panels correspond to perturbation widths $\sigma = 100$, $\sigma = 10$, $\sigma = 1$, and a single-wire $\delta$-function ($\sigma \to 0$).

For the broadest perturbation considered, $\sigma = 100 = \beta/\lambda$ -- a width that matches the softening length itself (Figure~\ref{fig:propagation}a) -- the initial perturbation behaves precisely as one might anticipate from textbook secular theory. The angular momentum deficit deposited at $j_{c}$ propagates coherently outward through the chain, generating a smoothly-evolving wave-like structure whose pattern marches across the annulus over the course of a few secular periods. As the width of the imposed perturbation is reduced, however, the morphology of the response changes qualitatively. In Figure~\ref{fig:propagation}(b), a perturbation of intermediate width ($\sigma = 10$) retains a clearly identifiable central peak in $\mathrm{Re}(z)$ throughout the integration window, while only a fraction of the imposed AMD bleeds outward into the surrounding wires (with a corresponding small contribution to the chain-edge response, where reflection from the finite boundary becomes visible). By $\sigma = 1$ (Figure~\ref{fig:propagation}c) the perturbation is essentially trapped at its launching site: a tall, persistent peak occupies the immediate neighborhood of $j_{c}$, and only a faint wisp of $\mathrm{Re}(z)$ and $\mathrm{Im}(z)$ leaks into the surrounding chain. The trend culminates in the delta-function limit (Figure~\ref{fig:propagation}d), in which a single excited wire generates a spike that is essentially stationary on the chain over the full ten secular periods displayed -- the AMD imposed at $j_{c}$ does not propagate appreciably outward over the integration window. We have confirmed that this behavior is not merely a short-time transient by extending the integration to $100\,T_{\rm sec}$, over which the localized morphology is preserved (see also Appendix~\ref{app:B}).

This behavior is, at first sight, surprising. The bare Laplace--Lagrange coupling is long-ranged: every wire of the chain is gravitationally coupled to every other through the matrix $\mathbf{A}$, and the short-wavelength limit -- where the perturbation is concentrated on a single wire or a handful of wires -- is precisely the regime in which the coupling between the excited wire and its immediate neighbors is strongest. One might therefore naively expect a delta-function perturbation to be the most efficiently \emph{conducting} configuration of all: an injection of AMD at a single location should excite nearby wires within a single secular time and redistribute the perturbation across the chain. Figure~\ref{fig:propagation} exhibits exactly the opposite trend: the more spatially concentrated the perturbation, the more reluctant it is to propagate. What is the physical origin of this quasi-localization -- and which length scale, exactly, controls it?

\subsection{The Dispersion Relation of the Softened Chain}\label{sec:dispersion}

To answer this question, consider an idealized homogeneous analogue of the chain of Section~\ref{sec:setup_numerical}: a system of equal-mass wires, $m_{j} = \bar{m}$, sharing a common mean motion $\bar{n}$, whose couplings depend on the inter-wire separation $d = |j - k|$ alone. In this limit, the Laplace--Lagrange matrix acquires constant diagonals,
\begin{align}
A_{j,\,j\pm d} \;&=\; a_{d} \;=\; \nu\,C_{2}\bigl(\alpha_{d},\beta\bigr), \nonumber \\
A_{jj} \;&=\; a_{0} \;=\; 2\,\nu\sum_{d = 1}^{\infty} C_{1}\bigl(\alpha_{d},\beta\bigr),
\label{eqn:kernel}
\end{align}
where $\nu \equiv \bar{n}\,\bar{m}/\M$ and $\alpha_{d} = (1+\lambda)^{-d}$, and the equations of motion~(\ref{eqn:LLeom}) are solved exactly by plane waves. Substituting $z_{j} = \exp[\,i\,(k\,j + \omega\,t)\,]$, with $k \in [0, \pi]$ the wavenumber in radians per wire, collapses the matrix problem to a single line:
\begin{align}
\omega(k) &\;=\; a_{0} \;+\; 2\sum_{d = 1}^{\infty} a_{d}\,\cos(k\,d), \nonumber \\
v_{g}(k) &\;\equiv\; -\,\frac{d\omega}{dk} \;=\; 2\sum_{d = 1}^{\infty} d\,a_{d}\,\sin(k\,d),
\label{eqn:dispersion}
\end{align}
the familiar tight-binding form, in which the dispersion relation is nothing more than the Fourier cosine series of a row of the coupling matrix. The endpoints of the band correspond to the two extreme apsidal configurations the chain can adopt: $k = 0$ to a chain-wide apsidally aligned state, and $k = \pi$ to alternating wire-to-wire anti-alignment. A localized wave packet of characteristic width $\sigma$ carries spectral weight at wavenumbers $k \lesssim 1/\sigma$, each component advancing at its own group velocity $v_{g}(k)$, the sign of which follows from the stationary-phase condition $dj/dt = -d\omega/dk$ for the ansatz adopted above (throughout, we refer to $k^{-1}$ as the spatial scale of a disturbance).

\begin{figure*}
\includegraphics[width=\textwidth]{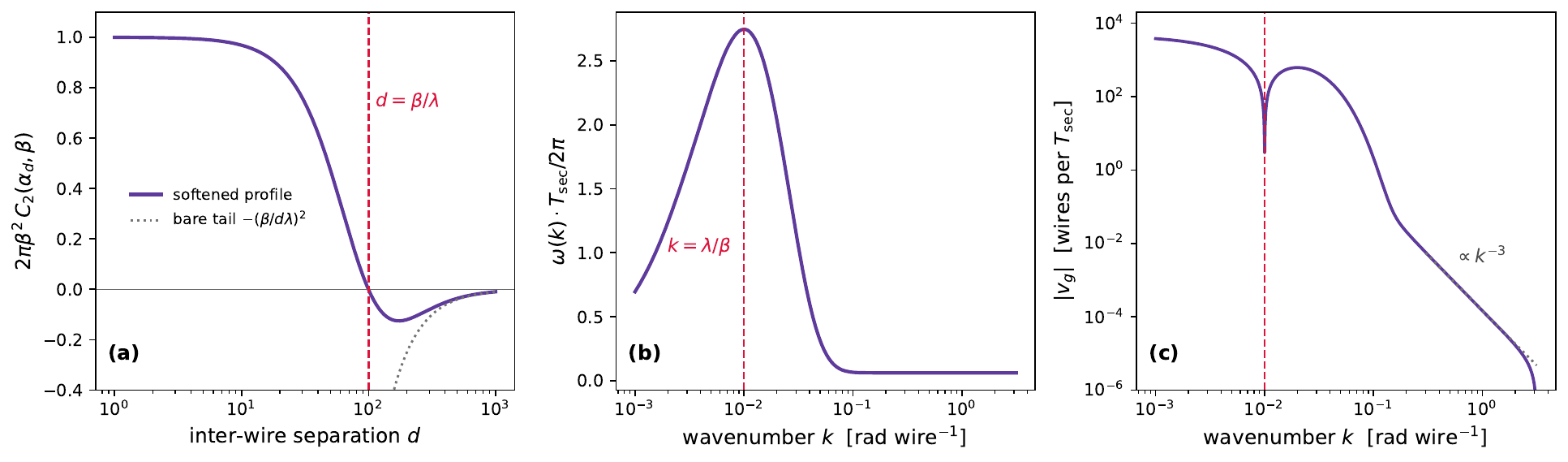}
\centering
\caption{The dispersion relation of the softened Laplace--Lagrange chain, in the infinite homogeneous idealization of equations~(\ref{eqn:kernel})--(\ref{eqn:dispersion}), with the Fourier sums carried to full convergence (see text). Panel~(a): the coupling profile $C_{2}(\alpha_{d},\beta)$, normalized by its dense-limit plateau value $1/(2\pi\beta^{2})$: flat within one softening length, sign reversal at $d = \beta/\lambda$ (dashed line), and reversion to the bare gravitational tail $-(\beta/d\lambda)^{2}$ (dotted curve) beyond. Panel~(b): the dispersion relation in apsidal cycles per $T_{\rm sec}$. All structure is confined to $k \lesssim$ a few $\lambda/\beta$: the precession rate peaks at $k \simeq \lambda/\beta$, where $v_{g}$ passes through zero (the notch in panel~c) before changing sign, and for $k > 0.5$ the band is flat to a few parts in $10^{5}$ of its span. (The uniform offset of $\omega$ -- the mean precession rate -- depends on the total mass within the coupled range and is dynamically inert; only the $k$-dependence of $\omega$ controls transport.) Panel~(c): the magnitude of the group velocity: the collective branch collapses quasi-exponentially below the softening scale, giving way for $k \gtrsim 0.3$ to the algebraic floor of equation~(\ref{eqn:floor}) (dotted line), imprinted by the residual $|x|$ cusp of the softened profile at small separations (Appendix~\ref{app:A}).}
\label{fig:dispersion}
\end{figure*}

The sums in equation~(\ref{eqn:dispersion}) are those of an infinite chain. We evaluate the couplings out to $d = 1.5\times10^{5}$, beyond which they are numerically negligible -- the pairwise coefficient decays exponentially once the fractional separation $d\lambda$ exceeds a few, so the sum terminates with no boundary discontinuity -- and verify the result against Abel-regularized partial sums at the $10^{-5}$ wires-per-$T_{\rm sec}$ level. (Naive truncation at $d \sim N$, by contrast, produces spurious boundary ringing in $v_{g}$ of amplitude $\sim\!3/k$ wires per $T_{\rm sec}$, entirely absent here.) Three related constructions appear in this paper, differing in their treatment of the boundaries and, in the case of the graded chain, in the radial variation of mass and mean motion: the infinite homogeneous chain, whose dispersion relation is shown in Figure~\ref{fig:dispersion}; the finite graded chain of Section~\ref{sec:setup_numerical}, on which the time evolutions of Section~\ref{sec:propagation} are carried out; and a finite periodic chain used below to propagate initial spectra exactly. For the periodic chain, each pairwise coupling is summed over all periodic images of the pair, $a^{\rm ring}_{d} = \sum_{m} a_{|d + mN|}$ with $d = |j - k|$, which is equivalent to sampling the dispersion relation of the infinite chain at the ring's wavenumbers $k_{n} = 2\pi n/N$; the simpler alternative of evaluating each coupling at the shorter index separation across the periodic boundary, $d_{jk} = \min(|j-k|,\,N-|j-k|)$, leaves a slope discontinuity in the coupling profile at $d = N/2$, whose consequences at the latest times we discuss below. We state which construction is in use wherever the distinction matters.

The structure of $\omega(k)$ is fixed entirely by the shape of the coupling profile $a_{d}$, shown in Figure~\ref{fig:dispersion}(a). The asymptotic analysis of Appendix~\ref{app:A} endows this profile with a simple form: within one softening length of the source -- separations $d \lesssim \beta/\lambda$ -- the coupling sits on a plateau of nearly uniform strength $+\nu/(2\pi\beta^{2})$, while beyond it the profile changes sign (numerically, at $d = 100$ wires for our parameters -- that is, at $d = \beta/\lambda$ on the nose) and decays along the weak unsoftened tail $-\nu/[2\pi\,(d\,\lambda)^{2}]$. In other words, once softening is accounted for, the dynamically dominant part of the secular interaction becomes effectively finite-ranged: a plateau of width $\beta/\lambda$, fringed by a much weaker algebraic tail of the opposite sign.

Two limits of the Fourier sum in equation~(\ref{eqn:dispersion}) then do all of the work. At $k \to 0$, the cosine factors add coherently, and $\omega(0)$ collapses to the row sum of the matrix -- where the near-cancellation $C_{1} \simeq -C_{2}$ derived in Appendix~\ref{app:A} annihilates the $\mathcal{O}(1/\beta^{2})$ plateau contributions and leaves only a logarithmic residue. The cancellation has a simple meaning. When $C_{1} = -C_{2}$ exactly, the pairwise secular energy in equation~(\ref{eqn:Hsec}) reduces to $\mathcal{H}_{jk} \propto C_{1}\,|z_{j} - z_{k}|^{2}$, a function of the relative displacement of the two eccentricity vectors alone, as if the wires were joined by springs that resist relative displacement but are indifferent to a common shift. A chain-wide translation of the eccentricity vectors ($k = 0$) stretches no spring, and the aligned end of the band retains only the sum over all separations of the small residual precession, of order $\nu\ln(1/\varepsilon)$ per pair, that Appendix~\ref{app:A} derives; the alternating configuration at $k = \pi$ stretches every spring joining wires an odd number of sites apart maximally, while wires an even number of sites apart move together. The aligned/anti-aligned structure itself is not a consequence of softening -- $C_{1} \simeq -C_{2}$ holds in the unsoftened limit too, and for a closely spaced pair it is the familiar decomposition of the secular dynamics into a conserved center-of-mass eccentricity vector and a uniformly precessing relative one \citep{2024ApJ...968...20Y} -- but softening sets the magnitude and sign of the anti-aligned precession rate (Appendix~\ref{app:A}) and, more consequentially, the width of the plateau over which it is exerted. The collective structure of the band is thereby compressed into wavenumbers of order $\lambda/\beta$, where the precession rate peaks at $k \simeq \lambda/\beta$, reaching $\simeq 2.7$ apsidal cycles per $T_{\rm sec}$ for our parameters.

At the opposite extreme, $k \gg \lambda/\beta$, the factor $\cos(k\,d)$ executes many oscillations across the plateau, and the sum self-averages: the band is flat to a few parts in $10^{5}$ of its full span for $k > 0.5$, with every short-wavelength mode precessing at a nearly common rate. All of the dispersion in $\omega(k)$ -- and hence all of the chain's capacity to transport an eccentricity signal -- resides at wavenumbers of order $\lambda/\beta$, below a few times that value (Figure~\ref{fig:dispersion}b). We note that the continuum counterpart of this flattening -- the suppression of self-gravity at wavelengths below the disk thickness, which renders the precession rates of short slow modes nearly wavelength-independent -- is a familiar feature of the $m = 1$ dispersion relation for softened self-gravitating discs \citep{2001AJ....121.1776T, 2003ApJ...595..531H}, and is the direct analogue of the reduction factor by which finite velocity dispersion weakens the collective gravitational response of a warm stellar disc relative to its cold counterpart \citep{2008gady.book.....B}; the discrete chain realizes it in particularly transparent form.

The same result admits a simple physical interpretation. Softening introduces a finite spatial resolution into the secular interaction. Perturbations whose characteristic scales exceed the softening length remain dynamically distinguishable, and therefore precess at different rates -- leading to phase mixing and dispersion. Once the spatial scale falls below the softening length, however, the interaction can no longer effectively distinguish progressively shorter structures: all sufficiently narrow perturbations produce nearly the same local pattern of relative displacements, and therefore precess at nearly the same rate. The resulting suppression of differential precession prevents phase mixing, allowing the initial perturbation to remain coherent over extraordinarily long times.

The group velocity extracted from this curve (Figure~\ref{fig:dispersion}c) organizes the propagation documented in Figure~\ref{fig:propagation}. Its magnitude peaks at $\simeq 3.4\times10^{3}$ wires per $T_{\rm sec}$ for scales comparable to the softening length -- the fastest signals the chain supports, crossing the full $N = 1000$ annulus in $\sim\!0.3\,T_{\rm sec}$ -- and collapses quasi-exponentially toward smaller scales, with a measured $e$-folding scale of $\simeq 1.3\,\lambda/\beta$ in $k$. This is the discrete counterpart of the exponential factor $e^{-|q|b}$ that multiplies the self-gravity term in the WKB dispersion relation of a Plummer-softened disc, $\omega = \dot{\varpi} + (\pi \G \Sigma/\Omega)\,|q|\,e^{-|q|b}$ (\citealt{2001AJ....121.1776T}, eq.~14; see also \citealt{2003ApJ...595..531H}), where $q$ is the physical radial wavenumber and $b$ the softening length. With $q = k/(\lambda a)$ and $b = \beta a$, the exponent is $|q|b = k\beta/\lambda$, and the maximum of $\omega$ at $|q| = 1/b$ maps onto the peak at $k = \lambda/\beta$ noted above. The group velocity, being the derivative of the exponentially suppressed term, inherits the same $e$-folding scale; it passes through zero at the peak of $\omega$ and changes sign beyond it, with a weaker lobe of opposite sign that dies away by $k \sim$ several $\lambda/\beta$.

The collapse does not, however, proceed to zero. The softened interaction retains a residual non-smoothness at coincident guiding centers: expanding the pair coefficient at small fractional separation $x$ gives $C_{2}(x,\beta) = C_{2}(0,\beta) - \tfrac{5}{4\pi\beta^{2}}\,|x| + \mathcal{O}(x^{2})$, the linear term being precisely the subleading coefficient derived in Appendix~\ref{app:A}, evaluated in the $\lambda \ll \beta$ regime. Because the high-$k$ asymptotics of a Fourier series are controlled by the short-distance regularity of its kernel, this $|x|$ cusp imprints an algebraic floor on the group velocity,
\begin{align}
|v_{g}| \;&\simeq\; \frac{5\,\nu\,\lambda}{\pi\,\beta^{2}}\,k^{-3} \nonumber \\
\;&=\; 1.47\times10^{-4}\,\left(\frac{k}{{\rm rad\,wire}^{-1}}\right)^{-3} \;\frac{\rm wires}{T_{\rm sec}},
\label{eqn:floor}
\end{align}
an analytic prediction valid in the intermediate regime $\lambda/\beta \ll k \ll 1$, where it matches the converged numerical curve to better than $0.1\%$ (Figure~\ref{fig:dispersion}c); at the zone edge, lattice symmetry forces $v_{g}(\pi) = 0$ exactly, and the numerical curve turns down accordingly. At $k = 1$, a packet advances by one wire per $\sim\!7\times10^{3}\,T_{\rm sec}$. Transport at sub-softening scales is therefore suppressed algebraically rather than exponentially for the coupling profile adopted here -- and we emphasize that both the magnitude of the floor and its algebraic form are properties of the adopted regularization and of the construction of the chain: under the vertical-thickness prescription of \citet{2003ApJ...595..531H}, as formulated by \citet{2019MNRAS.489.4176S}, the short-wavelength band is flatter still and the confinement correspondingly longer-lived, so that the Plummer baseline adopted here is the conservative choice (Appendix~\ref{app:C}).

To convert the dispersion relation into a parameter-free interpretation of Figure~\ref{fig:propagation}, we evolve the wavenumber spectrum $\tilde{z}(k, 0)$ of each of the four initial conditions exactly through the band. The calculation is performed on a periodic homogeneous chain of $N = 8192$ wires, each carrying the mean wire mass of the graded chain, coupled through the image-summed couplings defined above (with the corresponding $C_{1}$ diagonal), so that the discrete Fourier transform diagonalizes the system exactly; each initial spectrum is advanced as $\tilde{z}(k, t) = \tilde{z}(k, 0)\,e^{i\omega(k)t}$, and the retained AMD is evaluated at $t = 10\,T_{\rm sec}$ in the same $\pm\beta/\lambda$ window ($w = 201$ wires) applied to the experiments. Because a finite chain retains a residual background even for a fully dispersed perturbation (the window captures a fraction $w/N$ of uniformly spread AMD), we quote both the raw retention $R$ and the background-corrected trapped fraction $f = (R - w/N)/(1 - w/N)$. The spectral calculation predicts $f = 0.97$, $0.91$, $0.23$, and $0.00$ ($R = 0.974$, $0.909$, $0.252$, $0.023$) for the single-wire, $\sigma = 1$, $\sigma = 10$, and $\sigma = 100$ initial conditions, respectively; the graded-chain experiments of Figure~\ref{fig:propagation} yield $f = 0.97$, $0.91$, $0.29$, and $0.00$ ($R = 0.978$, $0.924$, $0.433$, $0.199$, the larger raw values reflecting the smaller chain, $w/N = 0.201$).

The band structure of the homogeneous chain, with no adjustable parameters, thus reproduces the confinement hierarchy of all four panels to within six percentage points -- the largest residual arising for $\sigma = 10$, the crossover case, where the frequency grading and finite boundaries of the true chain contribute at the several-percent level.

\begin{figure}
\includegraphics[width=\columnwidth]{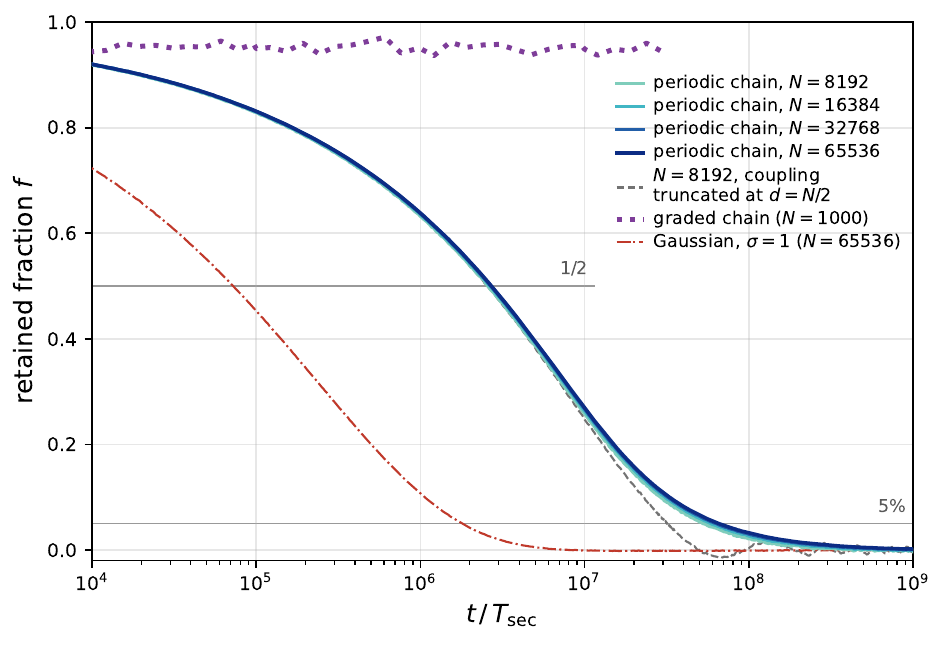}
\centering
\caption{Long-time retention of a single-wire perturbation on the periodic homogeneous chain. Solid curves: background-corrected retained fraction $f$ within the $\pm\beta/\lambda$ window for rings of $N = 8192$, $16\,384$, $32\,768$, and $65\,536$ wires built with the image-summed coupling, which converge on successive doublings of $N$. Gray dashed: the $N = 8192$ ring with the coupling truncated at $d = N/2$, which reproduces the half-life but empties the window about twice too early through the spurious antipodal coupling discussed in the text. Dotted purple: the graded chain of Section~\ref{sec:setup_numerical} ($N = 1000$), which plateaus and shows no measurable decay. Dash-dotted red: the $\sigma = 1$ Gaussian on the $N = 65\,536$ ring. Horizontal lines mark $f = 1/2$ and $f = 5\%$.}
\label{fig:retention}
\end{figure}

The same calculation, extended to long times, supplies a reference lifetime for the trapped core. The late-time evolution is sensitive to how the long-range coupling is treated on a finite ring. If the coupling is simply truncated at the periodic separation $d = N/2$, the slope discontinuity left in the coupling profile acts, on the ring's wavenumber grid, as a spurious coupling of each wire to its antipode; at $N = 8192$ this term exceeds the physical frequency spacing of adjacent modes by a factor of several for the slowest modes, $k \gtrsim 2.7$, which carry the late tail, and transfers them out of the retention window prematurely, while leaving the modes near $k \sim 1$ that set the half-life nearly untouched. The image-summed coupling is free of this defect, and with it the results converge on successive doublings of $N$ from $8192$ to $65\,536$ (Figure~\ref{fig:retention}): between the two highest resolutions, $32\,768$ and $65\,536$, the half-life changes by $0.6\%$ and the retained fraction at $3\times10^{7}\,T_{\rm sec}$ by $0.002$ (across the full range of $N$ the latter rises from $0.10$ to $0.11$). Carrying the exact spectral evolution to $10^{9}\,T_{\rm sec}$, we find that the single-wire core retains half of its AMD out to $t_{1/2} \simeq 2.7\times10^{6}\,T_{\rm sec}$, still retains $11\%$ at $3\times10^{7}\,T_{\rm sec}$, falls below a background-corrected retained fraction of $5\%$ by $\sim\!6.5\times10^{7}\,T_{\rm sec}$, and reaches $3\%$ at $10^{8}$ and zero to within $\pm 0.01$ at $10^{9}\,T_{\rm sec}$.

The decay is ballistic rather than diffusive: a mode of wavenumber $k$ exits the retention window once $|v_{g}(k)|\,t$ exceeds the window half-width, and this escape model, fed by the floor of equation~(\ref{eqn:floor}), reproduces the measured half-life to $3\%$; the slowest modes near the zone edge, where the lattice drives $v_{g} \to 0$, outlive the power law, and the late-time tail of the dispersal is correspondingly slower. The trapped core is thus not permanent, but it is extraordinarily long-lived -- its half-life exceeds the secular time by more than six orders of magnitude -- and the survival time of any given initial condition is set by its spectral weight at high $k$: the $\sigma = 1$ Gaussian, whose spectrum sits at smaller $k$ where the floor is higher, decays roughly forty times faster than the single wire ($t_{1/2} \simeq 7\times10^{4}\,T_{\rm sec}$). These figures pertain to the periodic homogeneous idealization, and they are conservative: evolving the actual graded chain of Section~\ref{sec:setup_numerical} over the same intervals (with the global precession phase removed to preserve numerical accuracy), the retained fraction plateaus near $0.95$ and exhibits no measurable decay through $3\times10^{7}\,T_{\rm sec}$ -- the frequency inhomogeneity of the real chain localizes its eigenmodes (Appendix~\ref{app:B}), suppressing even the residual ballistic leakage, so that the graded chain tested here confines the perturbation for longer than the homogeneous half-life.

In this language, the answer to the question posed at the close of Section~\ref{sec:propagation} takes its most intelligible form. The confinement documented there for $\sigma \ll \beta/\lambda$ is the statement that the perturbation's spectral content resides almost entirely on the flat portion of the band, where the modes' relative phases are nearly frozen and transport is suppressed by orders of magnitude; the crossover at $\sigma \sim \beta/\lambda$ occurs because $1/\sigma$ then passes the knee of $v_{g}(k)$, which sits at $k \sim \lambda/\beta$ because that is the reciprocal width of the coupling plateau. The apsidal polarization that surrounds a trapped wire is the real-space portrait of this spectral statement, as we show next (Section~\ref{sec:polarization}): the polarized configuration and the failure of the perturbation to propagate are not cause and effect, but two faces of the same nearly frozen superposition. The gradients of the true chain -- of mean motion, mass, and coupling strength across the annulus -- play no essential role in this account, a point we verify with a controlled experiment in Appendix~\ref{app:B}. The crossover is approximately universal: across chains spanning $\beta/\lambda = 20$--$200$, the retained fraction at $t = 10\,T_{\rm sec}$ within the $\pm\beta/\lambda$ window approximately collapses onto a single function of $\sigma\lambda/\beta$, with half of the imposed AMD retained at $\sigma\lambda/\beta \simeq 0.04$--$0.05$, and at fixed physical $\beta$ the trapped fraction is independent of the grid resolution (Appendix~\ref{app:C}).

The nearly flat short-wavelength portion of the secular band invites comparison with flat-band lattice systems studied in condensed-matter and photonic physics. In geometrically frustrated lattices such as the Lieb and kagome lattices, destructive interference can produce exactly dispersionless bands and \emph{compact localized states} -- finite-support superpositions of degenerate Bloch waves that do not diffract \citep{Leykam2018} -- and such states have been observed directly in photonic Lieb lattices \citep{Vicencio2015, Mukherjee2015}. The present chain is not an exact realization of this construction: it possesses a single band with a nearly flat short-wavelength sector, and its polarized excitation is neither strictly compact nor permanently stationary. Nevertheless, the same kinematic principle operates: the small frequency spread among the constituent modes strongly suppresses dephasing and transport, producing a long-lived, quasi-localized superposition, while the residual bandwidth ultimately allows it to disperse. Here, the band flattening arises not from lattice geometry, but from the softening-induced quenching of the collective gravitational response.

\subsection{Apsidal Polarization}\label{sec:polarization}

The confined response has a characteristic morphology, visible in Figure~\ref{fig:propagation}(c) and (d): the central peak is flanked by small negative excursions of the eccentricity vector in the frame co-rotating with the excited wire. These neighboring wires are mutually aligned but anti-aligned with respect to the center. We refer to this configuration as apsidally \emph{polarized}. Its structure follows directly from the spectral decomposition underlying the confinement described in Section~\ref{sec:dispersion}.

In the homogeneous chain, a single-wire perturbation has equal Fourier amplitude at every wavenumber. Its long-wavelength content occupies the dispersive portion of the band, while most of its AMD resides in the nearly flat short-wavelength sector. The escaping component has a smooth central profile extending over many wires. Once this component has dispersed from the launching region, the retained disturbance resembles the original spike minus that broad central profile: a narrow positive peak flanked by negative wings. The anti-alignment is therefore the spatial signature of removing the mobile part of the initial spectrum.

\begin{figure}
\includegraphics[width=\columnwidth]{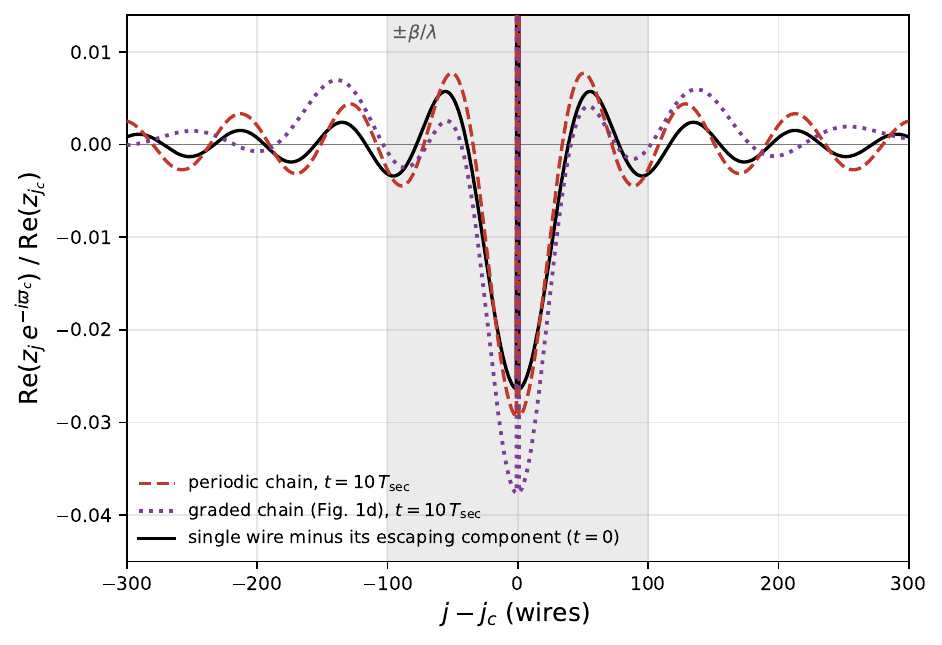}
\centering
\caption{The polarized morphology as a spectral projection. Solid black: the single-wire initial condition with its escaping long-wavelength content ($|k| < k_{c} = 8\,\lambda/\beta$) removed, at $t = 0$. Dashed red: the periodic homogeneous chain of Section~\ref{sec:dispersion} evolved to $t = 10\,T_{\rm sec}$. Dotted purple: the graded chain of Figure~\ref{fig:propagation}(d) at $t = 10\,T_{\rm sec}$. All profiles are de-rotated and normalized to the central wire, whose peak lies far above the top of the plot; the vertical scale is chosen to resolve the anti-aligned wings, and the shaded band marks the $\pm\beta/\lambda$ retention window.}
\label{fig:morphology}
\end{figure}

This interpretation can be tested directly. Removing from the single-wire initial condition the components with $|k| < k_{c}$ -- where the cutoff is fixed in advance from the dispersion relation as the largest wavenumber at which a mode still leaves the $\pm\beta/\lambda$ retention window within $10\,T_{\rm sec}$, that is, $|v_{g}(k_{c})|\,10\,T_{\rm sec} = \beta/\lambda$ evaluated on the outer flank beyond the second, opposite-sign lobe of $v_{g}$, giving $k_{c} \simeq 8\,\lambda/\beta$ -- removes $2.6\%$ of the initial AMD. The remaining profile already has negative wings with a depth of approximately $2.6\%$ of the central peak and a half-width of order $\pi/k_{c} \simeq 40$ wires. Figure~\ref{fig:morphology} compares this projected initial condition with the evolved profiles at $t = 10\,T_{\rm sec}$ on both the periodic homogeneous chain and the graded chain of Figure~\ref{fig:propagation}(d). The agreement in wing depth, width, and sign structure shows that the spectral projection accounts for the polarized morphology without requiring radial gradients. The precise cutoff, and hence the width of the wings, depends on the retention window and elapsed time.

Confinement and apsidal polarization thus follow from the same spectral structure: the mobile components disperse, leaving a slowly evolving short-wavelength superposition. The polarized profile is what a narrow perturbation looks like once its mobile part has left.

\medskip
\medskip
\medskip
\medskip
\medskip

\section{Conclusion}\label{sec:conclusion}

The analysis presented in this paper rests on a single observation. Whether or not a perturbation imposed on a softened Laplace--Lagrange chain propagates outward across the system depends critically on its wavelength. Long-wavelength perturbations evolve in the manner anticipated by textbook secular theory, dispersing coherently across the chain as the angular momentum deficit imposed at the launching site spreads outward through the surrounding wires. Short-wavelength perturbations, by contrast, do not propagate appreciably: over the integration windows considered here they remain trapped near their launching site, surrounded by an apsidally polarized shell of oppositely-aligned wires (Figure~\ref{fig:propagation}). Section~\ref{sec:dispersion} traces this dichotomy to the dispersion relation of the softened chain: the group velocity of eccentricity waves peaks for scales comparable to the softening length and collapses below it, so that a narrow perturbation is a superposition of modes with nearly frozen relative phases -- and the polarized configuration that surrounds it is the real-space portrait of that superposition. The crossover between the two regimes unfolds gradually as the perturbation width $\sigma$ is decreased; quasi-localization requires that the perturbation's spatial scale be much smaller than the softening length itself, which in the discrete chain corresponds to a width $\sigma \ll \beta/\lambda$ wires. For the parameter set adopted in this paper ($\beta = 10^{-2}$, $\lambda = 10^{-4}$), the softening length spans $\beta/\lambda = 100$ wires, the polarized shell that surrounds a trapped wire extends over about $40\%$ of that scale at $t = 10\,T_{\rm sec}$ -- a diagnostic scale, set by the wavenumber at which the group velocity collapses and by the elapsed time -- and confinement sets in for perturbations narrower still. At $t = 10\,T_{\rm sec}$ and within the $\pm\beta/\lambda$ window, half of the imposed AMD is retained for $\sigma \simeq 0.05\,\beta/\lambda$, consistent with the trend across Figure~\ref{fig:propagation}.

We propose that this behavior reflects the conceptual structure of the softening prescription itself. Softening is the procedure by which the inherent velocity dispersion of the constituent particles -- or, equivalently, the finite physical thickness of the disk -- is incorporated into the phase-averaged secular framework. Perturbations whose spatial extent is comparable to or smaller than the softening length therefore probe scales whose physical interpretation requires comparison with processes that the softened secular approximation omits.

A final remark concerns the relationship between the results presented here and the Schr\"odinger correspondence of the Laplace--Lagrange formalism \citep{2018MNRAS.475.5070B, MeltonBatygin}. The continuum reduction of the secular eccentricity dynamics onto a wave equation is, by construction, a long-wavelength approximation: it assumes that the complex eccentricity field varies smoothly on scales large compared with the softening length, and once spatial scales approach or fall below $\beta$, the derivative expansion underlying the wave equation ceases to be controlled. The flat-band quasi-localization identified in this work therefore marks the suppression of collective secular transport near the resolution scale of the softened theory. We emphasize that a real disk need not be dynamically inert on these scales: pressure, finite velocity dispersion, phase-dependent encounters, and other kinetic or collisional processes omitted from the present model may continue to redistribute AMD there. The quasi-localization thus delineates the short-wavelength boundary of the collective model, rather than an absolute barrier to transport.

For the concrete parameters adopted in this paper -- an annulus of roughly $1.5$ Mars masses near $1$\,au -- the homogeneous-chain half-life of a single-wire perturbation corresponds to $\gtrsim\!8\times10^{10}$\,yr, a figure that scales inversely with the annulus mass, and the graded chain confines it still longer: within softened secular theory, sufficiently narrow sub-softening AMD structure can outlive the system itself. The practical lesson for models of the annular architectures of Section 1 is accordingly one of caution: sub-softening structure in such calculations persists because the regularized collective theory transports it extremely slowly, and this by itself says nothing about whether such structure would be protected in nature.



\begin{acknowledgments}\nolinenumbers
We thank the anonymous referee for a constructive and careful report, and in particular for identifying the softening prescription to which our coefficients correspond. We are grateful to Tony Yap, Ian Brunton, Luke Handley, and Max Goldberg for insightful discussions. K.B. is grateful to the David and Lucile Packard Foundation, the Caltech Center for Comparative Planetary Evolution (3CPE), the National Science Foundation (grant number: AST 2408867), and NASA (Emerging Worlds grant number: 80NSSC26K0395) for their generous support.
\end{acknowledgments}

\software{{\tt numpy}, {\tt scipy}, {\tt matplotlib}}

\newpage

\appendix

\section{Verification of the Softened Coefficients against the Orbit-Averaged Disturbing Function}\label{app:verification}

To confirm that the prescription embodied in equations~(\ref{eqn:bsoft})--(\ref{eqn:C12}) faithfully represents a definite softened secular interaction, we benchmark it against a brute-force orbit-average of the regularized disturbing function. Given a pair of wires at semi-major axes $a < a'$, with eccentricities $(e, e')$ and apsidal longitudes $(\varpi, \varpi')$, we evaluate the doubly-averaged interaction
\begin{align}
\langle \mathcal{R}\rangle
&= \frac{1}{(2\pi)^{2}}\oint\!\!\oint
\frac{\G\,m'\,dM\,dM'}
{\sqrt{\,\mathcal{D}^{2}+\beta^{2}\,r'^{2}\,}},
\label{eqn:Ravg}
\end{align}
where $\mathcal{D}(M, M'; a, a', e, e', \varpi-\varpi')$ is the instantaneous in-plane separation between the two wires and $r'$ is the instantaneous radius of the outer wire, both expressed as functions of the mean anomalies $M$ and $M'$ via the standard Keplerian relations. The $\alpha$-derivative construction of equation~(\ref{eqn:C12}) rests on the fact that the pairwise potential, like the bare $1/|\mathbf{r}-\mathbf{r}'|$, is homogeneous of degree $-1$ in the two orbital radii, so that its dependence on $(a, a')$ enters only through $\alpha = a/a'$ and an overall factor of $1/a'$. A softening term must scale with the radii in the same way if the construction is to remain exact, and among such choices, $\beta^{2}r'^{2}$ -- a softening length that is a fixed fraction of the outer wire's instantaneous radius -- is the one for which the softened integrand reduces to $1 + \alpha^{2} + \beta^{2} - 2\alpha\cos\psi$, that is, to the coefficients $B^{(m)}_{1/2}(\alpha,\beta)$ of equation~(\ref{eqn:bsoft}). For radially non-overlapping orbits, it coincides with the $\beta^{2}\max(r^{2}, r'^{2})$ form tabulated by \citet{2019MNRAS.489.4176S} (their Table~B1). The comparison below tests the $r'^{2}$ form; at the smallest $\lambda$ of the $\beta = 100\lambda$ family the finite-difference eccentricity exceeds the fractional separation and the sampled orbits overlap radially, so a direct validation of the $\max$ form itself would require $\epsilon \ll \lambda$ there. A constant softening length, $\beta^{2}a'^{2}$, defines a different prescription. It agrees with equation~(\ref{eqn:Ravg}) for circular orbits, but its eccentricity derivatives differ, and the cross-coupling coefficient it generates departs from $C_{2}$ once $\beta$ approaches unity: by $12\%$, $84\%$, and $610\%$ at $\beta = 0.3$, $1$, and $3$ along the $\beta = 100\lambda$ family considered below. The $e^{2}$ self-coefficient, extracted at $e' = 0$ where $r' = a'$ identically, is insensitive to the distinction.

The integrals in equation~(\ref{eqn:Ravg}) are evaluated by direct summation on a uniform grid in the eccentric anomalies, with the Jacobian factors $(1 - e\cos E)\,(1 - e'\cos E')$ supplying the transformation of the mean-anomaly measure. Working in units in which $\G = m' = a' = 1$, the orbit-averaged disturbing function admits the small-eccentricity expansion
\begin{align}
\langle \mathcal{R}\rangle(e,\,e',\,\Delta\varpi)
= R_{0} + A\,\frac{e^{2}}{2} + A'\,\frac{e'^{2}}{2} + B\,e\,e'\cos\Delta\varpi + \ldots,
\label{eqn:Rexp}
\end{align}
where $\Delta\varpi \equiv \varpi - \varpi'$, $R_{0}$ is the circular background, and the ellipsis stands for higher powers of the eccentricities. Consistency of the softened prescription requires $A = A' = C_{1}$ and $B = C_{2}$; the two self-coefficients coincide because the homogeneity noted above makes the outer wire's expansion the mirror image of the inner wire's, exactly as in the unsoftened theory \citep{1999ssd..book.....M}. Choosing a finite-difference step $\epsilon = \min(10^{-3}, \beta/30)$ in eccentricity -- small enough that the $\mathcal{O}(\epsilon^{2}/\beta^{2})$ truncation error of the finite differences is at most $\sim\!10^{-3}$ at every point of the sweep, and large enough to avoid loss of precision through cancellation -- the three coefficients are isolated by centered differences of $\langle\mathcal{R}\rangle$ at a small set of sample points,
\begin{align}
A_{\rm dir} &= \frac{\langle\mathcal{R}\rangle(+\epsilon,0,0) + \langle\mathcal{R}\rangle(-\epsilon,0,0) - 2\,\langle\mathcal{R}\rangle(0,0,0)}{\epsilon^{2}}, \nonumber \\
A'_{\rm dir} &= \frac{\langle\mathcal{R}\rangle(0,+\epsilon,0) + \langle\mathcal{R}\rangle(0,-\epsilon,0) - 2\,\langle\mathcal{R}\rangle(0,0,0)}{\epsilon^{2}}, \nonumber \\
B_{\rm dir} &= \frac{\langle\mathcal{R}\rangle(\epsilon,\,\epsilon,\,0) - \langle\mathcal{R}\rangle(\epsilon,\,\epsilon,\,\pi)}{2\,\epsilon^{2}},
\label{eqn:stencils}
\end{align}
in which a negative eccentricity denotes an apsidal line advanced by $\pi$. The centered second differences cancel the background $R_{0}$ and return the self-coefficients up to $\mathcal{O}(\epsilon^{2})$ truncation, while the $\pi$ shift in $\Delta\varpi$ flips the sign of the cross-harmonic alone.

\begin{figure*}
\includegraphics[width=\textwidth]{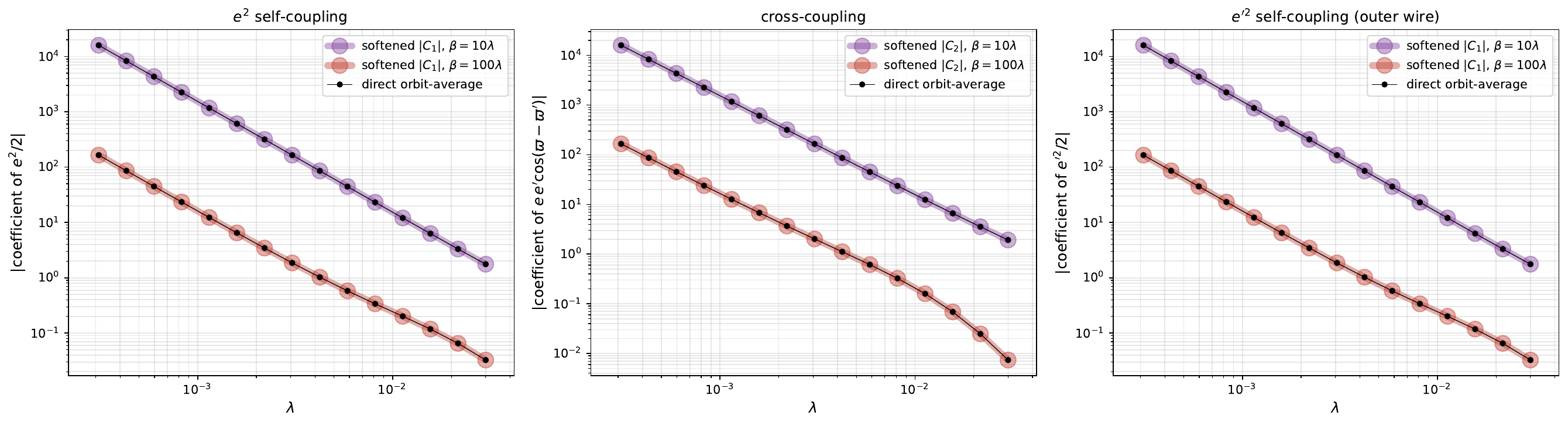}
\centering
\caption{Verification of the softened coefficients against a direct orbit-average of the softened disturbing function, equation~(\ref{eqn:Ravg}). Thick semi-transparent purple curves: the analytic forms of equation~(\ref{eqn:C12}) at $\beta = 10\lambda$; thick semi-transparent red curves: the same quantities at $\beta = 100\lambda$. Thin black lines with dotted markers: the coefficients extracted from the brute-force orbit-average through the finite-difference combinations of equation~(\ref{eqn:stencils}). Left: the inner wire's $e^{2}$ self-coupling, compared with $C_{1}$; center: the $e\,e'\cos(\varpi-\varpi')$ cross-coupling, compared with $C_{2}$; right: the outer wire's $e'^{2}$ self-coupling, also compared with $C_{1}$. The analytic and direct evaluations agree to plotting precision over the full two decades in $\lambda$ and both softening families, including where $\beta$ exceeds unity. Absolute values are plotted; both coefficients are sign-reversed relative to their unsoftened counterparts throughout the range shown.}
\label{fig:validation}
\end{figure*}

We carry out the comparison along two one-parameter families, $\alpha = 1/(1+\lambda)$ with $\beta = 10\lambda$ and $\beta = 100\lambda$, sweeping $\lambda$ logarithmically over two decades from $3\times10^{-4}$ to $3\times10^{-2}$, on a $10^{4}\times10^{4}$ grid in the eccentric anomalies that samples the closest-approach peak of the integrand by at least $30$ points along its narrowest direction at every point of the sweep. The results are shown in Figure~\ref{fig:validation}. Across the full sweep, the direct orbit-average and the analytic softened form agree in all three coefficients to better than $1.6\times10^{-3}$ at every point (and to better than $4\times10^{-4}$ in the two self-coefficients), the residual being the truncation error of the finite differences; the agreement reaches the $10^{-5}$ level for $\lambda \gtrsim 3\times10^{-3}$, including the points at $\beta > 1$ where the constant-softening prescription departs from $C_{2}$ by factors of several. The outer wire's $e'^{2}$ self-coefficient coincides with $C_{1}$ to the same precision as the inner wire's, as the homogeneity of the potential requires (under a constant softening length, by contrast, the two self-coefficients differ by up to $12\%$ over the same sweep). In the classical limit $\beta \to 0$ (tested at $\lambda = 0.3$, $\beta = 10^{-6}$), all three stencils reduce to the textbook Laplace--Lagrange values $\tfrac{1}{4}\alpha\,b^{(1)}_{3/2}$ and $-\tfrac{1}{4}\alpha\,b^{(2)}_{3/2}$ to better than $3\times10^{-5}$. Softening reverses the signs of both coefficients relative to their unsoftened counterparts over the range shown ($C_{1} < 0$, $C_{2} > 0$), and the figure plots their absolute values.

\section{Asymptotic Expansion of $C_{1}$ and $C_{2}$ for Small $\lambda$ and $\beta$}\label{app:A}
\setcounter{equation}{0}

In this Appendix we derive the leading- and logarithmic-order asymptotic behaviors of the softened secular coefficients $C_{1}$ and $C_{2}$ in the joint limit $\lambda \ll 1$ and $\beta \ll 1$, with the natural regulator scale
\begin{align}
\varepsilon^{2} \;\equiv\; \lambda^{2} + \beta^{2}.
\end{align}
The calculation proceeds by a matched-asymptotic argument: the orbit-averaged integrands defining $C_{1}$ and $C_{2}$ are sharply peaked at $\psi = 0$ when $\varepsilon \ll 1$, and we evaluate the local contribution exactly while leaving the (analytic, regular) outer contribution implicit through a matching constant. An alternative route to the same expansions is available through the closed-form expressions for the softened Laplace coefficients in terms of complete elliptic integrals (\citealt{2019MNRAS.489.4176S}, Appendix~C3), whose expansions as the modulus approaches unity are standard. We retain the matched-asymptotic route because it exposes the origin of the singular and logarithmic terms individually, and because the subleading coefficient it produces is the one that controls the transport floor of Section~\ref{sec:dispersion}.

\subsection*{Starting integrands}

Direct differentiation of the softened Laplace coefficient (\ref{eqn:bsoft}) yields the coefficients of equation~(\ref{eqn:C12}) in the explicit form
\begin{align}
C_{1} = \frac{1}{2\pi}\!\int_{0}^{\pi}\!\! \Bigl[\, & -\alpha\,u\,D^{-3/2} + \frac{3}{4}\,\alpha^{2}\,u^{2}\,D^{-5/2} \nonumber \\
& - \alpha^{2}\,D^{-3/2}\,\Bigr]\,d\psi,
\label{eqn:C1_explicit}
\end{align}
\begin{align}
C_{2} = \frac{1}{2\pi}\!\int_{0}^{\pi}\!\! \cos\psi\,\Bigl[\,& 2\,D^{-1/2} + \alpha\,u\,D^{-3/2} \nonumber \\
& - \frac{3}{4}\,\alpha^{2}\,u^{2}\,D^{-5/2} + \alpha^{2}\,D^{-3/2}\,\Bigr]\,d\psi,
\label{eqn:C2_explicit}
\end{align}
where $D \equiv 1+\alpha^{2}+\beta^{2}-2\alpha\cos\psi$ and $u \equiv 2\alpha-2\cos\psi$.

\subsection*{Local approximation}

Substituting $\alpha = 1 - \lambda$ into the definitions of $D$ and $u$\footnote{Throughout this appendix, $\lambda$ denotes the fractional separation $1 - \alpha$ of the pair. For nearest neighbors of the geometric chain of Section~\ref{sec:setup_numerical}, $1 - \alpha = \lambda/(1+\lambda)$, so the identification modifies the retained coefficients only at relative order $\lambda$ -- negligible for the parameters adopted in this paper.} gives
\begin{align}
D &= \lambda^{2}+\beta^{2} + 4(1-\lambda)\sin^{2}(\psi/2), \nonumber \\
u &= 4\sin^{2}(\psi/2) - 2\lambda.
\end{align}
Expanding $\sin^{2}(\psi/2) = \psi^{2}/4 - \psi^{4}/48 + \ldots\,$ in the small-$\psi$ region where the integrands peak, we obtain
\begin{align}
D \;\simeq\; \psi^{2} + \varepsilon^{2} - \frac{1}{12}\psi^{4} + \mathcal{O}(\lambda\psi^{2}, \psi^{6}), \nonumber \\
u \;\simeq\; \psi^{2} - 2\lambda + \mathcal{O}(\psi^{4}, \lambda\psi^{2}).
\end{align}
We work in the local region $|\psi| < A$ with $\varepsilon \ll A \ll 1$; the matching constant $A$ enters every logarithmic integral as $\ln(2A/\varepsilon)$ and cancels against the outer contribution, leaving only the $\ln\varepsilon^{2} = \ln(\lambda^{2}+\beta^{2})$ pieces in the final result. We retain the leading $D \simeq \psi^{2}+\varepsilon^{2}$ together with the $-\psi^{4}/12$ correction: as we shall see, the latter contributes at the logarithmic order we are computing through its effect on the $\alpha^{2}D^{-3/2}$ pieces of the two integrands.

\subsection*{Basic local integrals}

We define two families of local integrals on $[0,A]$. The first are convergent in the $A\to\infty$ limit and supply the power-law $\varepsilon^{-n}$ pieces:
\begin{align}
I_{a} &\equiv \int_{0}^{A}\!\!\frac{d\psi}{(\psi^{2}+\varepsilon^{2})^{3/2}} \;\simeq\; \frac{1}{\varepsilon^{2}}, \nonumber \\
I_{b} &\equiv \int_{0}^{A}\!\!\frac{d\psi}{(\psi^{2}+\varepsilon^{2})^{5/2}} \;\simeq\; \frac{2}{3\,\varepsilon^{4}}, \nonumber \\
I_{c} &\equiv \int_{0}^{A}\!\!\frac{\psi^{2}\,d\psi}{(\psi^{2}+\varepsilon^{2})^{5/2}} \;\simeq\; \frac{1}{3\,\varepsilon^{2}}.
\end{align}
The second are logarithmically divergent in $A$ and supply the $\ln\varepsilon^{2}$ pieces:
\begin{align}
L_{1} &\equiv \int_{0}^{A}\!\!\frac{d\psi}{(\psi^{2}+\varepsilon^{2})^{1/2}} \;\simeq\; \ln(2A/\varepsilon), \nonumber \\
L_{3} &\equiv \int_{0}^{A}\!\!\frac{\psi^{2}\,d\psi}{(\psi^{2}+\varepsilon^{2})^{3/2}} \;\simeq\; \ln(2A/\varepsilon) - 1, \nonumber \\
L_{5} &\equiv \int_{0}^{A}\!\!\frac{\psi^{4}\,d\psi}{(\psi^{2}+\varepsilon^{2})^{5/2}} \;\simeq\; \ln(2A/\varepsilon) - \frac{4}{3}.
\end{align}
The third of these follows from $\psi^{4} = (\psi^{2}+\varepsilon^{2})^{2} - 2\varepsilon^{2}(\psi^{2}+\varepsilon^{2}) + \varepsilon^{4}$, which reduces $L_{5}$ to $L_{1}$, $I_{a}$, and $I_{b}$.

\subsection*{$C_{1}$ to logarithmic order}

Substituting the local approximations into equation~(\ref{eqn:C1_explicit}) and expanding $(\psi^{2}-2\lambda)^{2} = \psi^{4} - 4\lambda\psi^{2} + 4\lambda^{2}$, the six monomials of the integrand contribute, term by term:
\begin{equation}
\begin{aligned}
&-\psi^{2}/D^{3/2} \!&&\to\!&& -L_{3}, \\
&+2\lambda/D^{3/2} \!&&\to\!&& +2\lambda\,I_{a}, \\
&+\frac{3}{4}\psi^{4}/D^{5/2} \!&&\to\!&& +\frac{3}{4}\,L_{5}, \\
&-3\lambda\psi^{2}/D^{5/2} \!&&\to\!&& -3\lambda\,I_{c}, \\
&+3\lambda^{2}/D^{5/2} \!&&\to\!&& +3\lambda^{2}\,I_{b}, \\
&-1/D^{3/2} \!&&\to\!&& -I_{a}.
\end{aligned}
\label{eqn:C1_pieces}
\end{equation}
The first and third entries are logarithmic, with coefficients $-1$ and $+3/4$ on $\ln(2A/\varepsilon)$, while the remaining four are power-law. \emph{A further logarithmic contribution} comes from the $-\psi^{4}/12$ correction to $D$ in the last entry, $-\alpha^{2}/D^{3/2}$, which is not negligible at this order. Expanding
\begin{align}
D^{-3/2} \;=\; D_{0}^{-3/2}\,\bigl[1 + (1/8)\,\psi^{4}/D_{0} + \ldots \bigr],
\qquad D_{0} \equiv \psi^{2}+\varepsilon^{2},
\end{align}
and integrating term by term, the $-\alpha^{2}/D^{3/2}$ piece acquires the additional contribution
\begin{align}
-\,\frac{1}{8}\,L_{5} \;=\; -\frac{1}{8}\ln(2A/\varepsilon) + \frac{1}{6},
\end{align}
of the same logarithmic order as the leading pieces in equation~(\ref{eqn:C1_pieces}). The logarithmic coefficients therefore collect to
\begin{align}
-1 + \frac{3}{4} - \frac{1}{8} \;=\; -\frac{3}{8}
\quad\text{on}\quad \ln(2A/\varepsilon).
\end{align}
Before assembling the power-law pieces, we note that at the order of the subleading $\lambda/\varepsilon^{2}$ terms, the $\mathcal{O}(\lambda)$ structure of the integrand must be retained as well: the prefactors carry $\alpha = 1-\lambda$ and $\alpha^{2} = 1-2\lambda$, and $D$ acquires the correction $-\lambda\,\psi^{2}$, each of which feeds the power-law integrals at relative order $\lambda$ (for example, $-\alpha^{2}\,D^{-3/2} \rightarrow -(1-2\lambda)\,D_{0}^{-3/2}\,[\,1 + \tfrac{3}{2}\,\lambda\,\psi^{2}/D_{0} + \ldots\,]$ supplies the additional pieces $+2\lambda\,I_{a} - \tfrac{3}{2}\,\lambda\,I_{c}$). Collecting all power-law contributions using $I_{a}, I_{b}, I_{c}$ from above,
\begin{align}
-\,\frac{1}{\varepsilon^{2}} \;+\; \frac{5\,\lambda}{2\,\varepsilon^{2}} \;+\; \frac{2\,\lambda^{2}}{\varepsilon^{4}} \;-\; \frac{3\,\lambda^{3}}{\varepsilon^{4}},
\end{align}
one obtains, after dropping the $\ln(2A)$ matching constant and writing $\ln\varepsilon = \frac{1}{2}\ln(\lambda^{2}+\beta^{2})$,
\begin{align}
C_{1}(\lambda, \beta)
\;\simeq\;&\frac{\lambda^{2}-\beta^{2}}{2\pi\,(\lambda^{2}+\beta^{2})^{2}}
+ \frac{\lambda\,(5\,\beta^{2}-\lambda^{2})}{4\pi\,(\lambda^{2}+\beta^{2})^{2}} \nonumber \\
& + \frac{3}{32\pi}\,\ln(\lambda^{2}+\beta^{2}) \,+\, \mathrm{regular},
\label{eqn:C1_log_final}
\end{align}
where the leading $1/\varepsilon^{4}$ structure arises from rewriting $2\lambda^{2}/\varepsilon^{4} - 1/\varepsilon^{2}$ in terms of $\lambda^{2}-\beta^{2}=2\lambda^{2}-\varepsilon^{2}$, and the subleading term collects the $\mathcal{O}(\lambda)$ pieces through the identity $5\lambda/(2\varepsilon^{2}) - 3\lambda^{3}/\varepsilon^{4} = \lambda\,(5\,\beta^{2}-\lambda^{2})/(2\,\varepsilon^{4})$.

\subsection*{$C_{2}$ to logarithmic order}

The calculation for $C_{2}$ follows the same logic, with the additional complication that the $\cos\psi = 1 - \psi^{2}/2 + \mathcal{O}(\psi^{4})$ factor multiplies each piece. The leading log contributions from the four monomials of equation~(\ref{eqn:C2_explicit}) accumulate as follows:
\begin{itemize}
\item $2\cos\psi/D^{1/2}$: yields $+2L_{1}$, with $\psi^{2}\cos\psi$ corrections suppressed by $\varepsilon^{2}\ln\varepsilon$. Log coefficient $+2$.

\item $\cos\psi\cdot\alpha u / D^{3/2}$: at leading order in $\lambda$ and $\cos\psi$, $\int(\psi^{2}-2\lambda)/D^{3/2}\,d\psi = L_{3} - 2\lambda\,I_{a}$, with $\cos\psi$- and $\lambda$-suppressed log corrections. Log coefficient $+1$.

\item $-\frac{3}{4}\cos\psi\cdot\alpha^{2}u^{2}/D^{5/2}$: yields $-\frac{3}{4}L_{5} + 3\lambda\,I_{c} - 3\lambda^{2}\,I_{b}$ at leading order, plus regular $\psi^{6}/D^{5/2}$ and $\lambda$-suppressed pieces. Log coefficient $-3/4$.

\item $\cos\psi\cdot\alpha^{2}/D^{3/2}$: yields $I_{a} - \frac{1}{2}L_{3} + \frac{1}{8}L_{5}$, where the three terms come, respectively, from the leading $\cos\psi\to 1$ piece, from the $-\psi^{2}/2$ correction to $\cos\psi$, and from the $-\psi^{4}/12$ correction to $D$. Log coefficient $-\frac{1}{2} + \frac{1}{8} = -\frac{3}{8}$.
\end{itemize}
The sum of the four logarithmic coefficients is
\begin{align}
+2 + 1 - \frac{3}{4} - \frac{3}{8} \;=\; \frac{15}{8}
\quad\text{on}\quad \ln(2A/\varepsilon).
\end{align}
The power-law pieces aggregate, by the same routine as in $C_{1}$ -- including the $\mathcal{O}(\lambda)$ prefactor and $D$-corrections retained there -- into the exact mirror of the $C_{1}$ collection, giving
\begin{align}
C_{2}(\lambda, \beta)
\;\simeq\;&-\frac{\lambda^{2}-\beta^{2}}{2\pi\,(\lambda^{2}+\beta^{2})^{2}}
- \frac{\lambda\,(5\,\beta^{2}-\lambda^{2})}{4\pi\,(\lambda^{2}+\beta^{2})^{2}} \nonumber \\
& - \frac{15}{32\pi}\,\ln(\lambda^{2}+\beta^{2}) \,+\, \mathrm{regular}.
\label{eqn:C2_log_final}
\end{align}

\subsection*{Summary and physical content}

Combining equations~(\ref{eqn:C1_log_final}) and~(\ref{eqn:C2_log_final}), the sum and difference take the compact forms
\begin{align}
C_{1}+C_{2} &\;\simeq\; -\frac{3}{8\pi}\,\ln(\lambda^{2}+\beta^{2}) \,+\, \mathrm{regular}, \nonumber \\[4pt]
C_{1}-C_{2} &\;\simeq\; \frac{\lambda^{2}-\beta^{2}}{\pi\,(\lambda^{2}+\beta^{2})^{2}}
+ \frac{\lambda\,(5\,\beta^{2}-\lambda^{2})}{2\pi\,(\lambda^{2}+\beta^{2})^{2}} \nonumber \\
& + \frac{9}{16\pi}\,\ln(\lambda^{2}+\beta^{2}) \,+\, \mathrm{regular}.
\label{eqn:C12_combinations}
\end{align}
The leading $\varepsilon^{-4}$ and $\varepsilon^{-2}$ singularities of $C_{1}$ and $C_{2}$ are exactly antisymmetric and cancel identically in the sum $C_{1}+C_{2}$ that controls the precession rate of the chain-wide aligned configuration at the $k \to 0$ end of the band (Section~\ref{sec:dispersion}). The logarithmic pieces, by contrast, are \emph{not} antisymmetric: with coefficients $+3/(32\pi)$ and $-15/(32\pi)$, they reinforce in the sum to give a non-vanishing logarithmic eigenfrequency for the aligned mode. Each of the coefficients in equation~(\ref{eqn:C12_combinations}) is reproduced by direct numerical quadrature to better than a percent -- at $\alpha = 1$ for the logarithmic pieces, and along rays of fixed $\lambda/\beta$ for the power-law terms.

A few limiting cases are instructive. At $\beta = 0$ the leading piece gives $C_{1} \simeq +1/(2\pi\lambda^{2})$ and $C_{2} \simeq -1/(2\pi\lambda^{2})$: the unsoftened form recovers the familiar $(1-\alpha)^{-2}$ divergence at $\alpha \to 1$, with $\lambda$ now playing the role of the only regulator. At $\lambda = 0$ the leading piece reverses sign, giving $C_{1} \simeq -1/(2\pi\beta^{2})$ and $C_{2} \simeq +1/(2\pi\beta^{2})$. The transition between these two regimes occurs at $\lambda \sim \beta$; at exactly $\lambda = \beta$ the leading $\varepsilon^{-4}$ piece vanishes and the subleading term takes over, reducing to $\pm 1/(4\pi\lambda)$.

The dynamical content of equation~(\ref{eqn:C12_combinations}) is that the apsidally aligned configuration -- the $k \to 0$ end of the band of Section~\ref{sec:dispersion}, whose precession rate is set by $\nu(C_{1}+C_{2})$ for a single pair and by the sum of this quantity over all separations for the chain -- is \emph{not} an exact zero mode at finite softening: each pair contributes a small, non-vanishing precession rate of order $\nu\,\ln(1/\varepsilon)$, parametrically smaller than the anti-aligned rate $\nu(C_{1}-C_{2})$ by a factor of $\varepsilon^{2}\,\ln(1/\varepsilon)$. This is the hierarchy visible in Figure~\ref{fig:dispersion}(b): the structure of the band is concentrated at long wavelengths, while its aligned end retains only the logarithmic residue.

\section{Eigenstructure of the Graded Chain, van Kampen Modes, and a Homogeneous Control Experiment}\label{app:B}
\setcounter{equation}{0}

The dispersion relation of Section~\ref{sec:dispersion} accounts for the confinement of narrow perturbations in terms of the nearly flat short-wavelength sector of an idealized homogeneous chain. For completeness, we document here the eigenstructure of the full Laplace--Lagrange matrix $\mathbf{A}$ of the graded chain, relate its localized eigenmodes to the singular modes of the continuum problem, and present a control experiment that isolates the ingredients on which the quasi-localization actually depends. The frequency inhomogeneity of the graded chain localizes its eigenvectors, and we identify these localized eigenvectors with discretized van Kampen modes of the continuum problem, an identification supported numerically below. The confinement of a narrow perturbation does not require them -- a periodic homogeneous chain, whose eigenmodes are extended Bloch waves, confines just as well -- but the inhomogeneity plays a genuine secondary role in suppressing the slow long-time leakage that the homogeneous chain exhibits.

\begin{figure*}
\includegraphics[width=\textwidth]{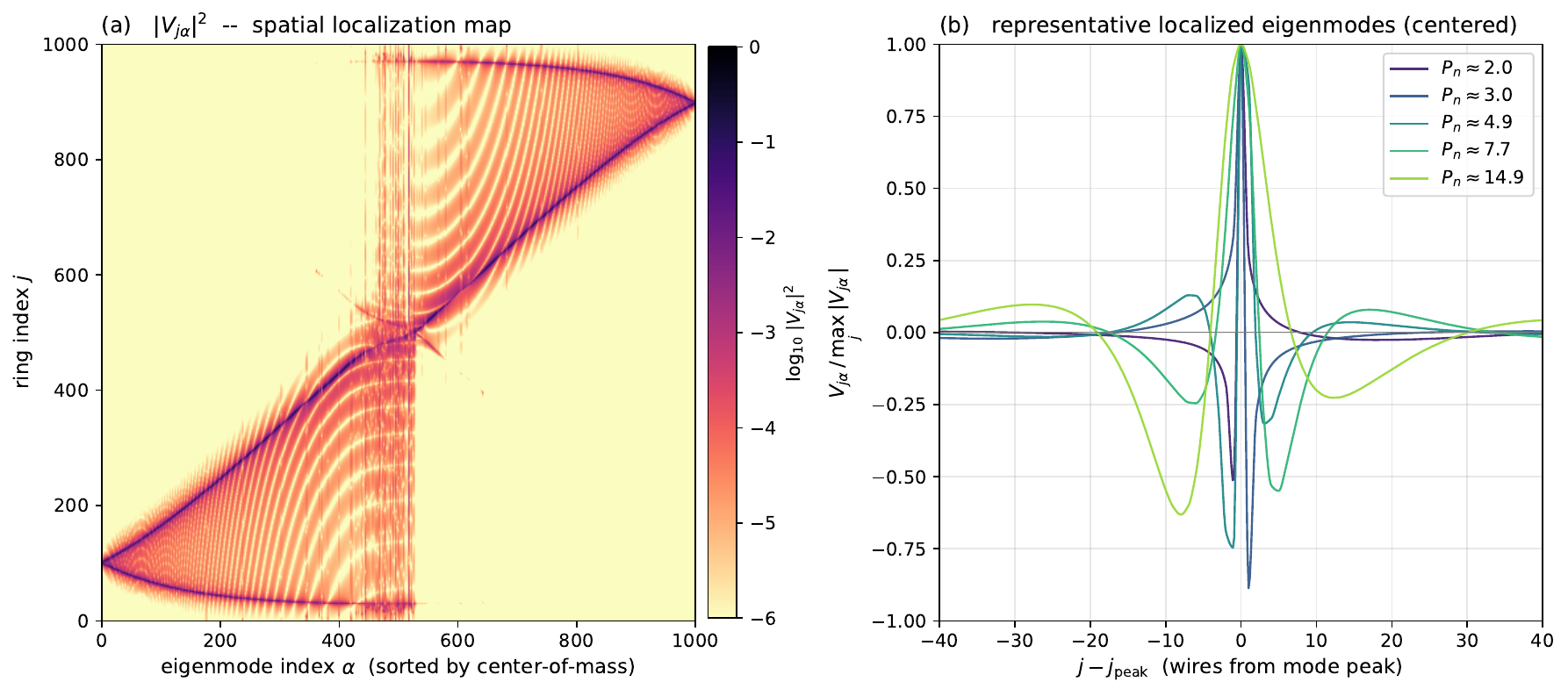}
\centering
\caption{Eigenstructure of the softened Laplace--Lagrange matrix $\mathbf{A}$ for the parameter set of Section~\ref{sec:setup_numerical} ($N = 1000$, $\lambda = 10^{-4}$, $\beta = 100\lambda$).  Panel (a): two-dimensional map of $|V_{j\alpha}|^{2}$ on a logarithmic color scale (high amplitude dark), with eigenmodes sorted by their spatial center of mass on the $\alpha$ axis and ring index $j$ on the vertical axis.  The dark diagonal stripe is formed by the compact eigenvectors of the dense band discussed in the text, each confined to a few wires about its center of mass.  Panel (b): five representative members of the dense band, each shifted so that its peak lies at the center of the plot and labeled by its participation ratio $P_{n}$ (equation~\ref{eqn:PR}).  All five exhibit the alternating-sign envelope characteristic of the short-wavelength ($k \gg \lambda/\beta$) end of the band.}
\label{fig:eigenstructure}
\end{figure*}

Diagonalizing $\mathbf{A}$ for the parameter set of Section~\ref{sec:setup_numerical} and sorting the eigenmodes by their spatial center of mass, $\bar{j}_{\alpha} = \sum_{j} j\,|V_{j\alpha}|^{2}$, yields the two-dimensional map shown in Figure~\ref{fig:eigenstructure}(a). Its dominant feature is a sharply-defined diagonal stripe of high-amplitude (dark) pixels. The stripe is formed by the modes of the \emph{dense band}: the roughly $90\%$ of the eigenvalues that crowd into the narrow frequency interval corresponding to the flat sector of Figure~\ref{fig:dispersion}(b), as distinct from the sparse, widely-spaced low-$|\lambda_{n}|$ eigenvalues that realize the dispersive long-wavelength branch. Each dense-band mode contributes a spatially narrow column to the map, supported only in a small radial neighborhood of its center of mass, with a local profile that alternates in sign from wire to wire, as befits the short-wavelength end of the band (Figure~\ref{fig:eigenstructure}b). Off the stripe, the few low-$|\lambda_{n}|$ modes appear as fully extended vertical bands. To quantify the spatial extent of a mode we use its participation ratio,
\begin{align}
P_{n} \;=\; \frac{\bigl(\sum_{j} V_{jn}^{2}\bigr)^{2}}{\sum_{j} V_{jn}^{4}},
\label{eqn:PR}
\end{align}
which equals $N$ for a mode spread uniformly over the chain and $1$ for one confined to a single wire. (We evaluate it on the eigenvectors of $\mathbf{A}$; using the canonical amplitudes $\sqrt{\Lambda_{j}}\,V_{jn}$ instead, which measure the distribution of a mode's AMD, changes the values only at the percent level, since $\Lambda_{j}$ varies by $16\%$ across the annulus.) The median participation ratio over the dense band is $P \simeq 2.2$ wires.

The compactness of these eigenvectors is genuine, but it is a property of the chain's frequency inhomogeneity rather than of the band itself. In the graded chain, the site frequencies $A_{jj}$ vary with radius through the gradients of mass and mean motion; in any finite chain with open boundaries, they vary additionally because the long-range $+1/[2\pi\,(d\lambda)^{2}]$ tail of $C_{1}$ is truncated at the chain ends. In either case, the variation across the chain exceeds the intrinsic width of the nearly degenerate band by more than five orders of magnitude. The radial variation of the apsidal precession rates therefore localizes the modes of this nearly degenerate band. Each mode is concentrated near a wire whose diagonal precession rate $A_{jj}$ closely matches its eigenfrequency, the two agreeing to better than $10^{-4}$ at the mode's amplitude peak. Diagonalizing the homogeneous control chain introduced below returns compact eigenvectors with median $P \simeq 4$ for this reason alone, whereas the same operator with its diagonal held constant at the bulk value -- the idealization of equation~(\ref{eqn:kernel}) -- yields extended, numerically sine-like modes with $P \simeq 2N/3$.

We identify these localized modes with the singular modes that \citet{2001AJ....121.1776T} encountered in the continuum slow-mode problem, whose numerical solution returns many singular (van Kampen) modes occupying the continuous portion of the eigenvalue spectrum. In the continuum, a mode of frequency $\omega$ is singular at the radius where $\omega$ coincides with the local precession rate, and the nearly degenerate band is a continuous spectrum; discretization replaces each singular mode by an eigenvector localized on the grid. The signature expected of this correspondence is that the width of the singular core is set by the grid rather than by the physics, and we have tested it by diagonalizing the graded chain at fixed physical softening ($\beta = 10^{-2}$, fixed annulus and total mass) for $N = 500$, $1000$, $2000$, and $4000$ (Table~\ref{tab:vankampen}). The participation ratio of the dense-band modes remains at $P \simeq 2.1$--$2.2$ wires throughout, so that the physical width of a mode's core shrinks in proportion to the wire spacing ($\propto N^{-1.03}$). Each mode also carries a low-amplitude smooth tail. To measure the extent of the full mode, core and tail together, we use its root-mean-square (RMS) extent, $w_{n} = [\sum_{j} |V_{jn}|^{2}\,(a_{j} - \bar{a}_{n})^{2}]^{1/2}$ with $\bar{a}_{n} = \sum_{j} |V_{jn}|^{2}\,a_{j}$ and $\sum_{j} |V_{jn}|^{2} = 1$ -- the standard deviation in semi-major axis of the mode's amplitude-squared distribution, a quantity dominated by the tail, whereas the participation ratio is dominated by the core. The median RMS extent over the dense band stays fixed at $0.0064\,a \simeq 0.63\,\beta a$, constant to $3\%$ across the four resolutions. The median spacing between adjacent dense-band eigenvalues falls as $N^{-1.10}$, while the frequency limits of the band are unchanged to four significant figures. A core of vanishing physical width, a smooth part of fixed physical extent, and a discretized continuous spectrum are what discretized singular continuum modes would produce, and we regard the identification as well supported, although we have not formulated the continuum eigenproblem and established the correspondence formally. The confined states of the graded-chain evolutions of Section~\ref{sec:propagation} are, in this sense, superpositions of these van Kampen-like modes; on the periodic homogeneous chain of Section~\ref{sec:dispersion}, the same confined states are superpositions of extended Bloch waves, which is the point of the control experiment below.

\begin{table}[t]
\centering
\caption{Resolution dependence of the dense-band eigenmodes of the graded chain at fixed physical softening ($\beta = 10^{-2}$; annulus $[0.95, 1.05]$; total mass $5\times10^{-7}\,\M$). $P$ is the median participation ratio over the dense band; the core width is $P\lambda$; the RMS extent is the standard deviation $w_{n}$ of the mode's amplitude-squared distribution in semi-major axis (defined in the text), which includes its smooth tail; the level spacing is the median spacing of adjacent dense-band eigenvalues, in units of the band's frequency span, which is the same to four figures at all $N$.}
\label{tab:vankampen}
\begin{tabular}{ccccc}
\hline\hline
$N$ & $P$ (wires) & core width$/\beta a$ & RMS extent$/\beta a$ & level spacing$/$span \\
\hline
500  & 2.19 & 0.044 & 0.65 & $3.8\times10^{-4}$ \\
1000 & 2.16 & 0.022 & 0.64 & $1.7\times10^{-4}$ \\
2000 & 2.09 & 0.010 & 0.63 & $7.9\times10^{-5}$ \\
4000 & 2.08 & 0.005 & 0.63 & $3.8\times10^{-5}$ \\
\hline
\end{tabular}
\end{table}

This identification also clarifies why eigenvector localization is not what confines a perturbation. To isolate what the confinement of Figure~\ref{fig:propagation} does and does not require, we construct an index-homogeneous control chain: $N = 1000$ wires of equal mass $\bar{m} = M_{\rm disk}/N$, sharing a common mean motion, with couplings given by the same softened coefficients as in equation~(\ref{eqn:kernel}) -- a function of inter-wire separation alone. This system possesses no radial grading of mass or mean motion; its only frequency inhomogeneity is the boundary-truncation potential discussed above. Evolving the same single-wire initial condition on both systems yields retention histories that are indistinguishable between the two: the fraction of the initial AMD retained within $\pm\beta/\lambda$ wires of the launching site is $98\%$ at $10\,T_{\rm sec}$, and $97\%$ at $100\,T_{\rm sec}$, in both cases. Decisively, the periodic homogeneous chain of Section~\ref{sec:dispersion}, whose eigenmodes are exactly extended Bloch waves with no localization whatsoever, traps the single-wire perturbation just as efficiently. The nearly flat band is therefore sufficient to produce long-lived confinement, even when the individual eigenmodes are spatially extended. In the graded chain, localization of the eigenmodes further suppresses the slow leakage found in the homogeneous calculation (Section~\ref{sec:dispersion}).

\section{Universality and Prescription Dependence of the Confinement}\label{app:C}
\setcounter{equation}{0}

The experiments of Section~\ref{sec:experiments} employ a single discretization, $\lambda = 10^{-4}$ and $\beta = 100\lambda$. In this Appendix, we verify that the confinement they document is controlled by the dimensionless ratio $\sigma\lambda/\beta$ -- rather than by any feature of that particular grid -- and that it survives the replacement of the softening prescription itself.

\begin{figure*}
\includegraphics[width=\textwidth]{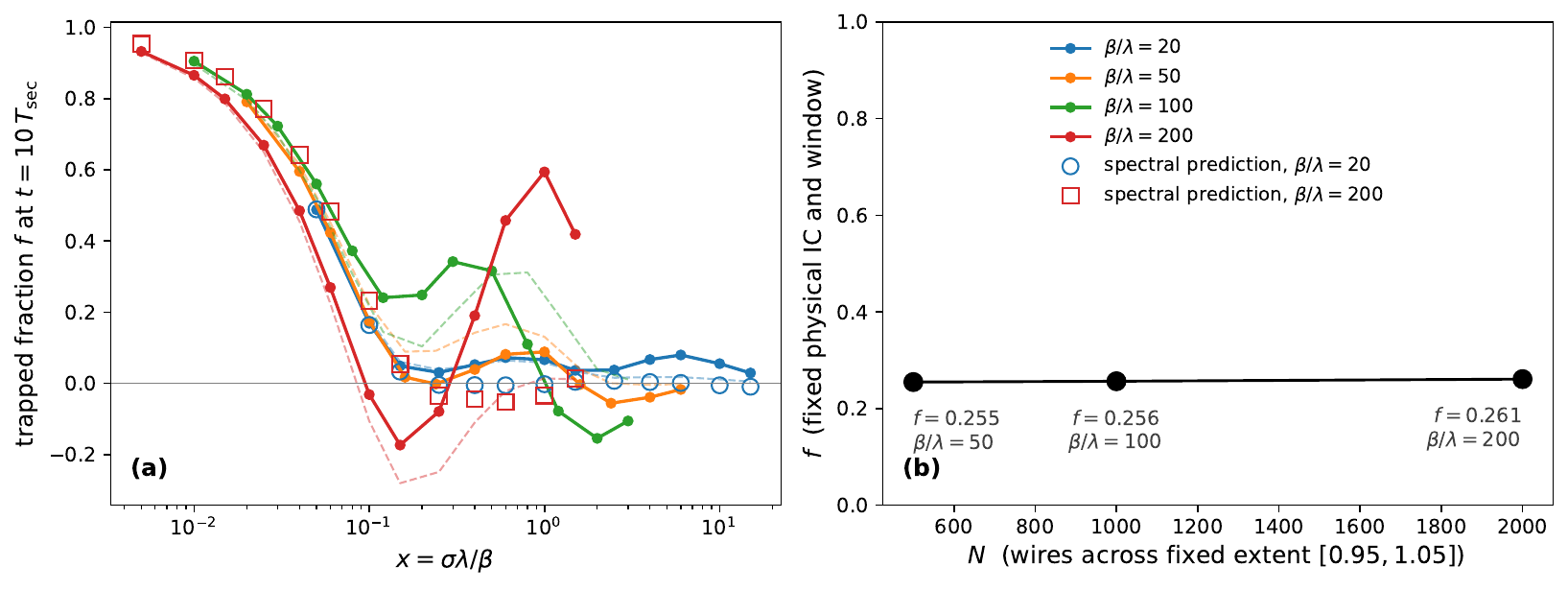}
\centering
\caption{Universality of the confinement. Panel~(a): background-corrected trapped fraction $f$ at $t = 10\,T_{\rm sec}$, measured within a window of half-width $\beta/\lambda$ wires, as a function of the rescaled perturbation width $x = \sigma\lambda/\beta$, for graded chains with $\beta/\lambda = 20$, $50$, $100$, and $200$ (solid curves: single snapshot; faint dashed curves: averaged over the slow beat of the finite chain). Open symbols give the parameter-free spectral predictions for the two extreme chains. The crossover approximately collapses onto a single function, with half-retention at $x \simeq 0.04$--$0.05$; the non-collapsing excursions at $x \gtrsim 0.3$ for the largest windows are a finite-size effect discussed in the text. Panel~(b): resolution study at fixed physical problem -- $\beta = 10^{-2}$, fixed radial extent, fixed physical perturbation width and retention window -- for $N = 500$, $1000$, and $2000$: the trapped fraction is resolution-independent to $2.4\%$.}
\label{fig:universality}
\end{figure*}

We first repeat the $\sigma$-sweep of Figure~\ref{fig:propagation} on graded chains with $\beta/\lambda = 20$, $50$, $100$, and $200$ (at fixed $\lambda = 10^{-4}$), measuring in each case the background-corrected trapped fraction $f$ within a window of half-width $\beta/\lambda$ wires at $t = 10\,T_{\rm sec}$, with $T_{\rm sec}$ evaluated per chain. Figure~\ref{fig:universality}(a) shows the four curves against the rescaled width $x = \sigma\lambda/\beta$, together with parameter-free spectral predictions of the type constructed in Section~\ref{sec:dispersion} for the two extreme cases. The crossover approximately collapses: half of the imposed AMD is retained at $x = 0.050$, $0.054$, $0.050$, and $0.037$ for the four chains, and the spread among the chains remains below $0.06$ throughout the crossover once the slow two-mode beat of the finite chain is averaged over. (For the widest case, $\beta/\lambda = 200$, the retention window spans $40\%$ of the chain, and the dispersed tail of the curve is contaminated by a coherent finite-size oscillation that is absent from the periodic spectral prediction; the collapse statement refers to the crossover itself, which is unaffected.)

A sharper statement follows from holding the physical problem fixed and varying only the resolution (Figure~\ref{fig:universality}b). Fixing $\beta = 10^{-2}$, the radial extent $[0.95, 1.05]$, a Gaussian perturbation of physical width $10^{-3}$, and a retention window of physical half-width $0.01 = \beta$, we vary $N \in \{500, 1000, 2000\}$, so that $\beta/\lambda \simeq 50$, $100$, and $200$. The trapped fraction at $10\,T_{\rm sec}$ is $f = 0.255$, $0.256$, and $0.261$: constant to $2.4\%$ across a factor of four in resolution. The confinement scale is the physical softening length, not the wire spacing -- the phenomenon is not an artifact of the discretization.

The long-time confinement, which is the quantity most sensitive to the residual dispersion of the band, can be examined in the same way, with one qualification. Evolving the single-wire perturbation on the same three chains by exact eigendecomposition, the retained fraction settles onto a flat plateau by $10^{2}\,T_{\rm sec}$ and remains there through $3\times10^{7}\,T_{\rm sec}$ at every resolution: over the interval $10^{6}$--$3\times10^{7}\,T_{\rm sec}$, the beat-averaged retained fraction varies by less than $0.02$, $0.01$, and $0.005$ about constant levels of $f \simeq 0.93$, $0.96$, and $0.98$ for $N = 500$, $1000$, and $2000$, with no trend, whereas the retained fraction of the periodic homogeneous chain, whose decay is ballistic with a slow late-time tail rather than exponential (Section~\ref{sec:dispersion}), has fallen to $\simeq 11\%$ by $3\times10^{7}\,T_{\rm sec}$ ($5\%$ by $\sim\!6.5\times10^{7}$ and $3\%$ by $10^{8}\,T_{\rm sec}$). The qualification is that a single excited wire is a grid-scale initial condition rather than a fixed physical one: the fraction of its AMD residing on the dispersive branch, which escapes at early times, is proportional to $\lambda/\beta$ (measured: $7.6\%$, $5.0\%$, and $2.4\%$), which is why the plateau level rises with $N$. The Gaussian of fixed physical width of panel~(b), followed to the same times, retains a residual core of $f \simeq 0.1$--$0.2$ at all three resolutions with no trend either, although the slow beat of the finite chain (amplitude $\pm 0.09$) limits how precisely its level can be compared across $N$. Because the secular clock itself is converged -- the median eigenperiod changes by $+0.3\%$ and $-0.2\%$ between $N = 1000$ and $N = 500$, $2000$ (the median single-wire precession period $2\pi/|A_{jj}|$ by $+2.3\%$ and $-1.2\%$) -- the same conclusions follow when the three chains are compared at fixed physical time rather than in units of their own $T_{\rm sec}$.

\begin{figure*}
\includegraphics[width=\textwidth]{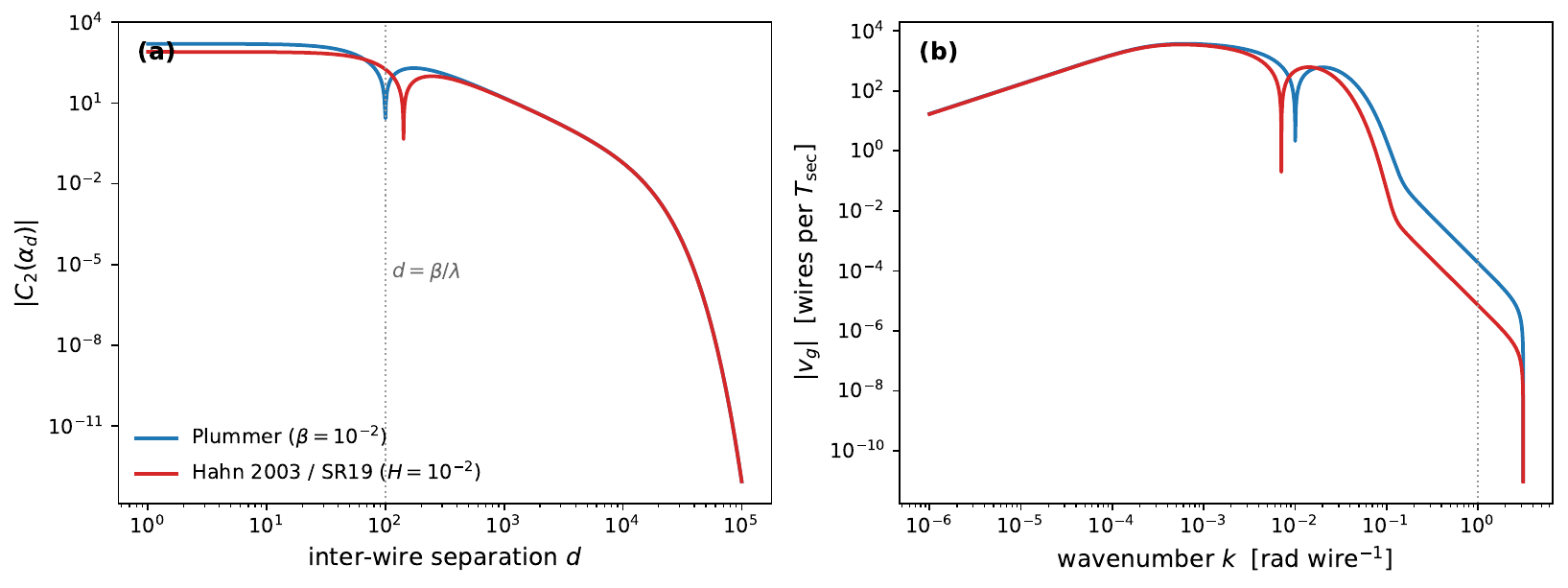}
\centering
\caption{Robustness to the softening prescription: the Plummer-in-the-integrand baseline of Section~\ref{sec:setup} (blue) versus the vertical-thickness prescription of \citet{2003ApJ...595..531H}, as formulated by \citet{2019MNRAS.489.4176S}, with $H = \beta$ (red). Panel~(a): magnitude of the ring--ring coupling coefficient $C_{2}$ versus separation. The two prescriptions differ only within the softening scale and join the bare gravitational interaction beyond it; the notch in each curve marks its change of sign. Panel~(b): group velocity of the homogeneous chain. The mobile long-wavelength branch is prescription-independent (peak speeds agree to $5\%$), and both prescriptions collapse to an algebraic floor scaling approximately as $k^{-3}$ at short wavelength, with the thickness prescription lying a factor of $\simeq 26$ lower at $k = 1$ for the AMD-symmetrized chain kernel defined in the text ($\simeq 13$ at matched effective near-neighbor softening; the ratio depends on the symmetrization weighting, see text).}
\label{fig:prescription}
\end{figure*}

Finally, because the coefficients of Section~\ref{sec:setup} embody one particular regularization, we repeat the analysis under the vertical-thickness prescription of \citet{2003ApJ...595..531H}, which \citet{2019MNRAS.489.4176S} identify as one of only two commonly used softening models that converge exactly to the razor-thin secular dynamics in the limit of vanishing softening, and which provides a physically motivated description of a vertically extended disc. Concretely, with scale height $H = \beta$, the coupling coefficients become $C_{1} = (\alpha/4)\,[\,B^{(1)}_{3/2} - 3\alpha H^{2}(2 + H^{2})\,B^{(0)}_{5/2}\,]$ and $C_{2} = -(\alpha/4)\,[\,B^{(2)}_{3/2} - 3\alpha H^{2}(2 + H^{2})\,B^{(1)}_{5/2}\,]$, where the $B^{(m)}_{s}$ are softened Laplace coefficients with softening $\varepsilon^{2}(\alpha) = H^{2}(1 + \alpha^{2})$ (eqs.~7 and 14, and Table~1, of \citealt{2019MNRAS.489.4176S}). Figure~\ref{fig:prescription} compares the two prescriptions. The coupling coefficients agree with the bare interaction beyond the softening scale and differ only within it (panel~a), and the dispersion relations are indistinguishable on the mobile long-wavelength branch, with peak group velocities agreeing to $5\%$ (panel~b). The short-wavelength sector flattens under both prescriptions -- to $2\times10^{-5}$ of the band span for the Plummer form, and a further factor of $\sim\!19$ flatter for the thickness prescription -- and the high-$k$ envelope of the group velocity is algebraic, $\propto k^{-3.15}$, in both comparison chains as constructed here. What depends on the prescription is the magnitude of the floor and, with it, the lifetime, and here the construction of the comparison chain must be stated, because the floor is set by the behavior of the chain's coupling at small separation. The homogeneous comparison chains of this Appendix are built from the AMD-symmetrized coupling of the graded chain. For wires of mass $m_{j} \propto a_{j}$, the two one-sided Laplace--Lagrange couplings of a pair at radius ratio $\alpha$ are $\nu\,\alpha^{3/2}\,C_{2}(\alpha)$ (outer wire forced by inner) and $\nu\,C_{2}(\alpha)$ (inner by outer), with $\nu = n\,m/M$ evaluated at the inner wire, and the similarity transformation $\sqrt{\Lambda_{k}/\Lambda_{j}}\,A_{kj}$ that renders $\mathbf{A}$ symmetric maps both onto their geometric mean, $a_{d} = \nu\,\alpha_{d}^{3/4}\,C_{2}(\alpha_{d})$, with $a_{0} = 2\nu\sum_{d}\alpha_{d}^{3/4}\,C_{1}(\alpha_{d})$ and the sums carried to $d = 10^{5}$; the ring coupling is summed over periodic images as in Section~\ref{sec:dispersion}. This kernel differs from the index-homogeneous kernel of the main text, $a_{d} = \nu\,C_{2}(\alpha_{d})$, only by the smooth factor $\alpha_{d}^{3/4}$. Under this construction, the thickness-prescription floor at $k = 1$ lies a factor of $\simeq 26$ below the Plummer one, and the half-life of a single-wire perturbation lengthens by the same factor ($5.4\times10^{7}$ versus $2.1\times10^{6}\,T_{\rm sec}$, converged on rings of up to $N = 65\,536$ wires; the latter differs from the $2.7\times10^{6}\,T_{\rm sec}$ of Section~\ref{sec:dispersion} through the $\alpha^{3/4}$ weighting). Part of this factor reflects the larger effective smoothing of the thickness regulator at like separations ($\varepsilon(\alpha{=}1) = \sqrt{2}\,H$): repeating the comparison at matched effective near-neighbor softening, $H = \beta/\sqrt{2}$, gives factors of $\simeq 13$ in both the floor and the half-life, with the floor exponent unchanged. The numerical factors quoted above depend on the radial weighting used to construct the comparison chain and are not universal. Both tested softening prescriptions nevertheless produce a nearly flat short-wavelength sector and long-lived confinement.

\end{document}